\documentclass[journal,twoside,final]{IEEEtran}

\usepackage{amsmath}
\usepackage{amssymb}
\usepackage{balance}
\usepackage{bm}
\usepackage{cite}
\usepackage{color}
\usepackage[T1]{fontenc}
\usepackage[cmintegrals]{newtxmath}
\usepackage{graphicx}
\usepackage[utf8]{inputenc}
\usepackage{longtable}
\usepackage{mathrsfs}
\usepackage{mathtools}
\usepackage{multirow}
\usepackage{float}
\usepackage{siunitx}
\usepackage{soul}               
\usepackage[color=green!40]{todonotes}
\soulregister\cite7
\soulregister\footnote7
\soulregister\eqref7
\soulregister\ref7
\usepackage{url}
\usepackage{threeparttable}
\usepackage{hyperref}
\newlength{\myvspace}

\graphicspath{{./}{figures/}}

\newcommand{\myauthor}{Xintao Huan, Kyeong Soo
  Kim,~\IEEEmembership{Senior~Member,~IEEE}, Sanghyuk Lee, Eng Gee
  Lim,~\IEEEmembership{Senior~Member,~IEEE}, and Alan
  Marshall,~\IEEEmembership{Senior~Member,~IEEE}}%

\newcommand{\mytitle}{A Beaconless Asymmetric Energy-Efficient Time
  Synchronization Scheme for Resource-Constrained Multi-Hop Wireless Sensor
  Networks}%

\begin{document}

\title{\LARGE \mytitle}

\author{\myauthor}%


\maketitle

\begin{abstract}
  The ever-increasing number of WSN deployments based on a large number of
  battery-powered, low-cost sensor nodes, which are limited in their computing
  and power resources, puts the focus of WSN time synchronization research on
  three major aspects, i.e., accuracy, energy consumption and computational
  complexity. In the literature, the latter two aspects haven't received much
  attention compared to the accuracy of WSN time synchronization. Especially in
  multi-hop WSNs, intermediate gateway nodes are overloaded with tasks for not
  only relaying messages but also a variety of computations for their offspring
  nodes as well as themselves. Therefore, not only minimizing the energy
  consumption but also lowering the computational complexity while maintaining
  the synchronization accuracy is crucial to the design of time synchronization
  schemes for resource-constrained sensor nodes. In this paper, focusing on the
  three aspects of WSN time synchronization, we introduce a framework of reverse
  asymmetric time synchronization for resource-constrained multi-hop WSNs and
  propose a beaconless energy-efficient time synchronization scheme based on
  reverse one-way message dissemination. Experimental results with a WSN testbed
  based on TelosB motes running TinyOS demonstrate that the proposed scheme
  conserves up to 95\% energy consumption compared to the flooding time
  synchronization protocol while achieving microsecond-level synchronization
  accuracy.
\end{abstract}

\begin{IEEEkeywords}
  Beaconless time synchronization, energy efficiency, reverse asymmetric time
  synchronization, wireless sensor networks.
\end{IEEEkeywords}

\IEEEpeerreviewmaketitle

\section{Introduction}
\label{sec:introduction}
\IEEEPARstart{T}{ime} synchronization for wireless sensor networks (WSNs) has
been extensively studied in the last decades as the number of WSN deployments
has been gradually increasing over the period \cite{sundararaman05:_clock,
  rhee09:_clock_synch_wirel_sensor_networ, Wu11:spm}.

Because many existing time synchronization schemes rely on sensor nodes'
broadcasting synchronization messages received from a root node (i.e.,
\textit{flooding}) to achieve network-level time synchronization (e.g.,
\cite{maroti04,kusy06:_elaps}), additional computation and energy consumption
are imposed on the resource-constrained sensor nodes which are already loaded
with tasks including medium access control (MAC) protocol, message scheduling
and routing, and data measurement.
The Flooding Time Synchronization Protocol (FTSP) \cite{maroti04}, which is a
benchmark among WSN time synchronization schemes, is a typical example of
flooding-based time synchronization schemes, where the energy consumption caused
by the layer-by-layer broadcasting and the computational complexity of
estimating the clock skew and offset could be high for battery-powered, low-cost
sensor nodes.


To address the high computational complexity of estimating the clock skew and
offset to achieve microsecond-level accuracy in FTSP, the Ratio-based time
Synchronization Protocol (RSP) has been proposed based on the periodical one-way
synchronization message dissemination and a lightweight procedure for the
estimation of clock parameters \cite{rsp}. In RSP, additional thresholds and
procedures are introduced to reduce the impact on the clock parameter estimation
of the numerical errors by the limited floating-point precision (i.e., 32-bit
floating-point numbers) of the resource-constrained sensor nodes. However, like
FTSP, RSP also relies on the periodical broadcasting of the synchronization
messages and, therefore, is not energy efficient in terms of message
transmissions by sensor nodes. Another well-known WSN time synchronization
scheme is the Timing-sync Protocol for Sensor Networks (TPSN)
\cite{ganeriwal03:_timin_protoc_sensor_networ}, which is based on the two-way
message exchange requiring additional response message along with the beacon
message. Considering that the schemes based on the two-way message exchange
could achieve higher accuracy in principle by compensating for propagation
delay, however, TPSN's synchronization error of tens of microseconds is quite
high even compared to that of the schemes based on one-way message
dissemination. In case of the Reference Broadcast Synchronization (RBS)
\cite{elson02:_fine}, because it requires not only additional reference nodes to
broadcast the common clock time but also additional message exchanges between
the sensor nodes to estimate their relative time differences, it is also not
energy-efficient and proper for resource-constrained sensor nodes. Aiming at
conserving the energy consumption through reducing message transmissions for the
synchronization procedures, the authors of \cite{EETS} proposed the
Energy-Efficient and rapid Time Synchronization (EETS). Although EETS could
reduce the message transmissions between neighboring levels in a hierarchical
network, it still strongly relies on the broadcasting of the synchronization
messages to achieve network-level time synchronization. Also, because the energy
efficiency of EETS is indirectly evaluated through the number of message
transmissions using simulations, its actual energy consumption could be
different on real sensor nodes.

The authors of \cite{Kim:17-1} proposed a novel energy-efficient time
synchronization scheme based on the reverse two-way message exchange and
demonstrated through simulation experiments that sub-microsecond-level
synchronization accuracy could be achieved. Still, the computational precision
required by their scheme (i.e., the precise division of the floating-point
numbers) is beyond the capability of most resource-constrained sensor nodes
equipped with a low-cost MicroController Unit (MCU) providing only 32-bit
floating-point as discovered in \cite{Huan:19-1}. The idea of the reverse
two-way message exchange together with message bundling for synchronization and
measurement data, however, could be applied for designing more advanced
energy-efficient time synchronization schemes targeting resource-constrained
sensor nodes because it can greatly reduce the number of message transmissions
by sensor nodes.

\begin{figure}[!tb]
  \begin{center}
    \includegraphics[width=\linewidth]{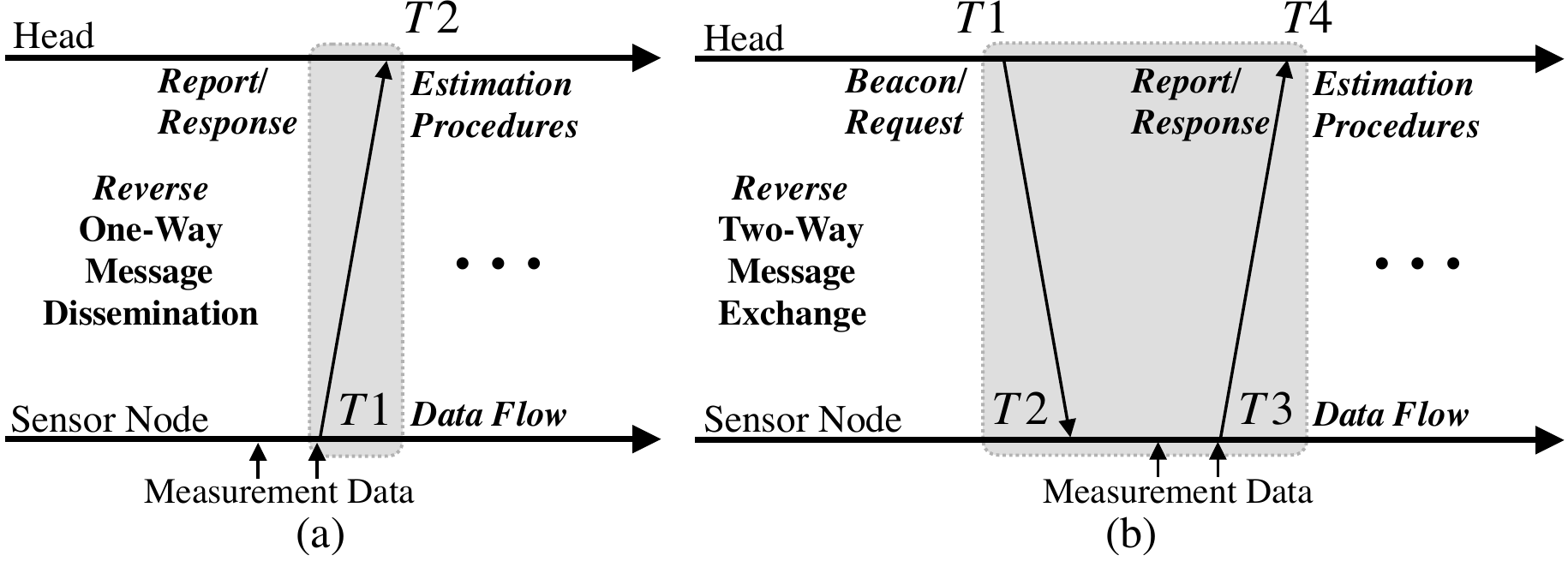}
  \end{center}
  \caption{Overview of the proposed reverse asymmetric time synchronization
    based on (a) one-way message dissemination and (b) two-way message exchange
    with optional bundling of measurement data.}
  \label{fig:reverse_oneandtwoway}
\end{figure}

In this paper, we present the reverse asymmetric time synchronization framework
illustrated in Fig.~\ref{fig:reverse_oneandtwoway} and propose the Beaconless
Asymmetric energy-efficient Time Synchronization (BATS) scheme specifically
based on the reverse one-way message dissemination shown in
Fig.~\ref{fig:reverse_oneandtwoway}~(a). In the proposed framework, all the
synchronization procedures are moved from sensor nodes to the head\footnote{The
  head node and the monitoring station (i.e., a PC or a workstation for data
  processing) connected to it locally or remotely over the Internet are
  collectively called the head in this paper.} as in \cite{Huan:19-1}, and
application messages (e.g., for reporting measurement data to the head) carry
synchronization-related data as well. In addition, BATS does not rely on the
``Beacon/Request'' messages as shown in Fig.~\ref{fig:reverse_oneandtwoway}~(a),
thereby saving energy for their transmissions and receptions at sensor nodes,
the latter of which often consume more energy \cite{telosb}. The proposed scheme
is designed to address the three major challenges in WSN time synchronization on
resource-constrained sensor nodes, i.e., achieving high synchronization
accuracy, reducing energy consumption \cite{CESP} and lowering computational
complexity \cite{3chars}.

Based on the reverse asymmetric time synchronization framework and the one-way
message dissemination, the proposed BATS scheme combines in a unique way the
advantages of several existing WSN time synchronization schemes: Being based on
the one-way message dissemination, BATS can take the advantage of simpler
implementation compared to those based on the two-way message exchange. Unlike
the conventional schemes based on the one-way message dissemination including
FTSP, however, BATS is optimized for energy efficiency by reversing the flow of
synchronization messages from ``head${\rightarrow}$sensor nodes'' to ``sensor
nodes${\rightarrow}$head'', which is the key idea of the reverse asymmetric time
synchronization framework. The reverse asymmetric time synchronization
framework, on the other hand, is a generalization of the schemes we have
proposed in \cite{Kim:17-1} and \cite{Huan:19-1} and belongs to a broader
category of \textit{reactive time synchronization protocols} \cite{sallai06},
where sensor nodes' local clocks, which are not synchronized to a common
reference, are used to timestamp events; synchronization takes place after the
event has been detected. However, BATS is also different from the existing
reactive time synchronization protocols like Routing Integrated Time
Synchronization protocol (RITS) \cite{kusy06:_elaps} in its multi-hop extension,
where, unlike RITS, there is no layer-by-layer time translation at gateway nodes
along a multi-hop path; instead, the said time translation is completely moved
to the head to relieve the burden of the gateway nodes.

Note that the actual energy consumption and synchronization accuracy of the
proposed scheme are evaluated and analyzed through experiments on a real WSN
testbed in this paper. To the best of our knowledge, this is the first work that
shows the actual performance of energy-efficiency of the high-precision time
synchronization schemes on resource-constrained sensor nodes through real
experiments.

The rest of the paper is organized as follows: In
Section~\ref{sec:related_work}, the related work including
the time synchronization scheme proposed in \cite{Kim:17-1} based on reverse
two-way message exchange and the conventional time synchronization schemes based
on one-way message dissemination is
discussed. Section~\ref{sec:energy-efficient_ts} presents the proposed BATS
scheme based on the reverse one-way message
dissemination. Section~\ref{sec:multi-hop extension} discusses the multi-hop
extension of BATS scheme with a focus on the issue of communication
overhead. The results of the performance evaluation of BATS scheme in terms of
energy consumption and synchronization accuracy on a real testbed are presented
in Section~\ref{sec:exp_results}. Section~\ref{sec:concluding-remarks} concludes
our work in this paper and highlights the future work.

\section{Related Work}
\label{sec:related_work}

\subsection{Time Synchronization Based on Reverse Two-Way Message Exchange}
\label{sec:reversetwoway}
Aiming at providing high-precision time synchronization for an asymmetric WSN
with a head equipped with abundant computing and power resources and multiple
resource-constrained sensor nodes (e.g., powered by standalone batteries and
equipped with low-cost MCUs), the authors of \cite{Kim:17-1} proposed an
energy-efficient time synchronization scheme based on the asynchronous source
clock frequency recovery (SCFR) and the reverse two-way message exchange shown
in Fig.~\ref{fig:reverse_oneandtwoway}~(b), which we call it EE-ASCFR in short
from now on. In EE-ASCFR, the estimation of the clock offset and frequency ratio
is separated into the head and the sensor node to reduce the computational
complexity of the latter where a logical clock is maintained based on the
estimated clock frequency ratio. Although the broadcasting of the beacon
messages is still required, the number of message transmissions at sensor nodes
is reduced through bundling several measurement data together with
synchronization data in a ``Report/Response'' message. Compared to BATS, parts
of the time synchronization procedures---i.e., the estimation of the clock
frequency ratio and the maintenance of the logical clock based on it---are still
done at sensor nodes in EE-ASCFR.

Since the hardware clocks of sensor nodes based on low-cost crystal oscillators
run independently at different clock frequencies and offsets with respect to
those of the reference clock, which is the hardware clock of the head in
\cite{Kim:17-1}, the hardware clock $T_{i}$ of a sensor node $i$ for a $N$-node
WSN (i.e., $i{\in}\left[0,1,\ldots,N{-}1\right]$) can be described by the affine
clock model as follows:
\begin{equation}
  \label{eq:hardware_clock_model}
  T_{i}(t) = \left(1+\epsilon_{i}\right)t + \theta_{i} ,
\end{equation}
where $t$ is the reference clock time, and
$\left(1{+}\epsilon_{i}\right){\in}\mathbb{R}_{+}$ and
$\theta_{i}{\in}\mathbb{R}$ are the clock frequency ratio and the clock offset,
respectively. The aforementioned logical clock
$\mathscr{T}^{\textnormal{EE-ASCFR}}_{i}$ maintained at the sensor node $i$ is
described in terms of the hardware clock $T_{i}$ as follows \cite{Kim:17-1}: For
$t_{k}{<}t{\leq}t_{k+1}$ ($k{=}0,1,\ldots$),
\begin{equation}
  \label{eq:ee-ascfr_logical_clock}
  \mathscr{T}^{\textnormal{EE-ASCFR}}_{i}\Big(T_{i}(t)\Big) = \mathscr{T}^{\textnormal{EE-ASCFR}}_{i}\Big(T_{i}(t_{k})\Big)
  + \dfrac{T_{i}(t)-T_{i}(t_{k})}{1 + \hat{\epsilon}_{i,k}} - \hat{\theta}_{i,k} ,
\end{equation}
where $t_{k}$ is the reference time for the $k$th synchronization, and
$\hat{\epsilon}_{i,k}$ and $\hat{\theta}_{i,k}$ are the clock skew and offset of
the sensor node $i$ estimated during the $k$th synchronization. Note that, since
the estimation of the offset and frequency ratio is separately done at the head
node and the sensor node in EE-ASCFR, the $\hat{\theta}_{i,k}$ is set to 0 in
\eqref{eq:ee-ascfr_logical_clock} for the logical clock at the sensor node, which
means that the logical clock is synchronized to the reference clock in frequency
but runs with a different offset.


As for the frequency ratio $1{+}\epsilon_{k}$ in
\eqref{eq:hardware_clock_model}, due to the use of the cumulative ratio (CR)
estimator of the asynchronous SCFR scheme proposed in \cite{Kim:13-1} and the
reverse two-way message exchange, it is estimated as
$(T2{-}T2_{init}){/}(T1{-}T1_{init})$ where $T1_{init}$ and $T2_{init}$ are the
timestamps recorded during the initial beacon message transmission at the head
and its reception at the sensor node, respectively, and $T1$ and $T2$ are the
timestamps corresponding to the latest beacon message. The time points of
recording the timestamps $T1$ and $T2$ are shown in
Fig.~\ref{fig:reverse_oneandtwoway}~(b).

Of particular note is that the second term in
\eqref{eq:ee-ascfr_logical_clock}---i.e.,
$\left(T_{i}(t)-T_{i}(t_{k})\right){/}\left(1 +
  \hat{\epsilon}_{i,k}\right)$---for the logical clock update requires the
division by the frequency ratio, instead of multiplication, unlike the time
synchronization schemes based on conventional two-way message
exchange. Considering the microsecond-level timestamps involved in the
calculation, this division may cause significant numerical computation error due
to the limited floating-point precision in the resource-constrained sensor nodes
embedded with low-cost MCU as discussed in \cite{rsp} and \cite{Huan:19-1}.


\subsection{Time Synchronization Based on One-Way Message Dissemination}
\label{sec:oneway}
\begin{figure}[!tb]
  \begin{center}
    \includegraphics[width=\linewidth]{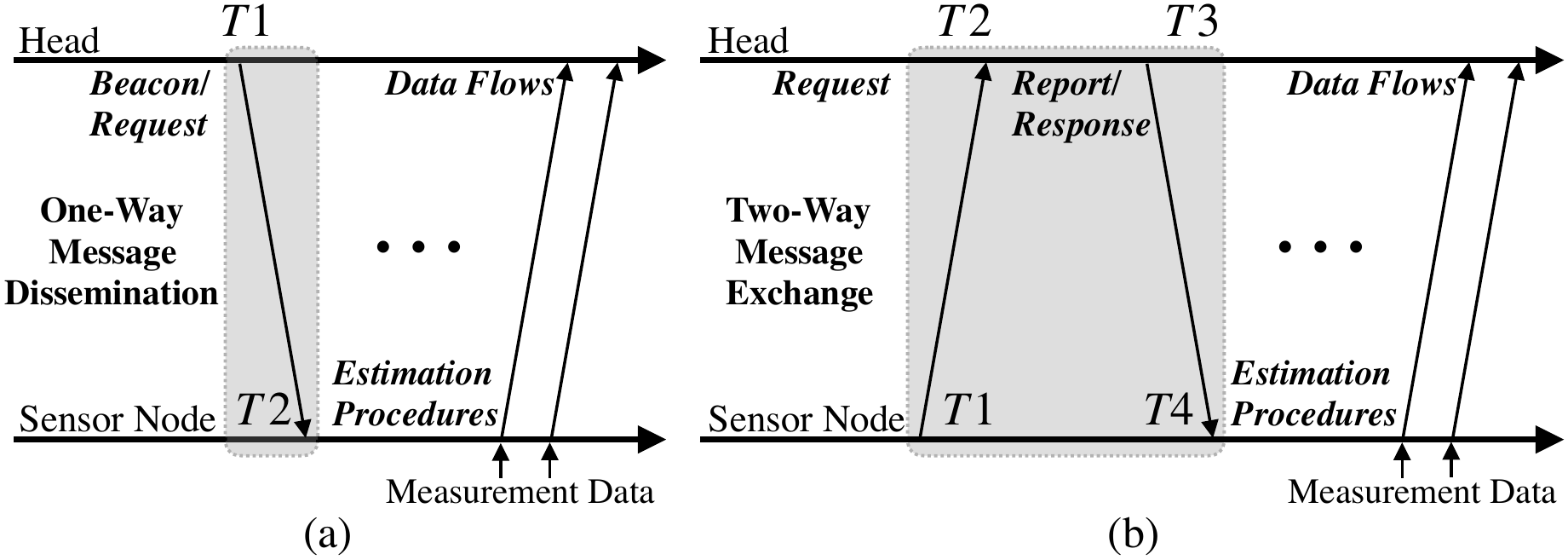}
  \end{center}
  \caption{Overview of the conventional time synchronization based on (a)
    one-way message dissemination and (b) two-way message exchange.}
  \label{fig:conventional_oneandtwoway}
\end{figure}

Compared to the time synchronization schemes based on the two-way message
exchange, those based on the one-way message dissemination have simpler
synchronization procedures as shown in
Fig.~\ref{fig:conventional_oneandtwoway}~(a) (i.e., conventional one) and
Fig.~\ref{fig:reverse_oneandtwoway}~(a) (i.e., reverse one). As a representative
example of the time synchronization schemes based on the one-way message
dissemination, here we discuss the details of FTSP,
whose implementation is available as part of the standard TinyOS library
\cite{tinyos}, and identify issues in its implementation on resource-constrained
WSN sensor nodes.

\begin{figure}[!tb]
  \centering \includegraphics[width=\linewidth]{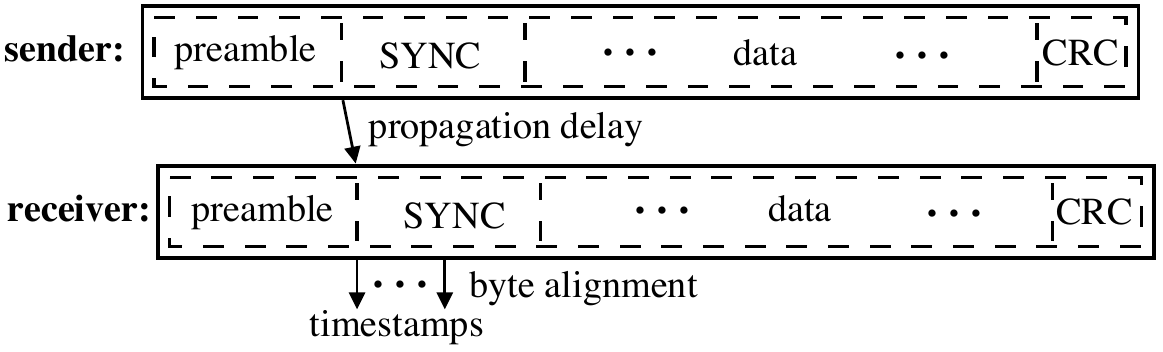}
  \caption{MAC-layer timestamping procedure introduced in FTSP \cite{maroti04}.}
  \label{fig:ftsp_timestamping}
\end{figure}

In FTSP, the linear regression is used to estimate the clock frequency ratio and
offset, i.e., the slope and the intercept of the regression line. However, due
to the limited computing resources of the sensor node, FTSP limits the number of
timestamp samples used for the least squares estimation of the slope and
intercept: Specifically, past 8 timestamp values are employed for all different
synchronization intervals (SIs) during the performance evaluation in
\cite{maroti04}. In addition to the linear-least-squares-based clock frequency
ratio and offset estimation, the timestamping procedure of FTSP records multiple
timestamps---i.e., timestamps at each byte boundary after the SYNC bytes as
shown in Fig.~\ref{fig:ftsp_timestamping}---in sending and receiving a
synchronization message to reduce the jitter of the interrupt handling and
encoding/decoding times, which is also quite demanding for resource-constrained
sensor nodes \cite{maroti04}.

The authors of \cite{rsp} proposed RSP, which is also based on the flooding of
synchronization messages but simplifies the estimation of the clock skew and
offset. In case of the head and the sensor node shown in
Fig.~\ref{fig:conventional_oneandtwoway}~(a), their hardware clock times (i.e.,
$t$ for the head and $T_{i}(t)$ for the sensor node) modeled in
\eqref{eq:rsp_hardware_clock} can be described as follows:
\begin{equation}
  \label{eq:rsp_hardware_clock}
  t = \alpha_{i}T_{i}(t) + \beta_{i} ,
\end{equation}
where $\alpha_{i}$ and $\beta_{i}$ are the clock frequency ratio and the clock
offset of the reference clock with respect to the sensor node hardware clock,
respectively.
Let $T1_{k}$ and $T2_{k}$ ($k{=}1,\ldots$) be the timestamps corresponding to
the latest Beacon/Request message received at the sensor node. Then, the logical
clock $\mathscr{T}^{\textnormal{RSP}}_{i}$ at the sensor node $i$---i.e., the
estimated reference time corresponding to the sensor node's hardware clock---is
obtained as follows \cite{rsp}:
\begin{equation}
  \label{eq:rsp_logical_clock}
  \mathscr{T}^{\textnormal{RSP}}_{i}\Big(T_{i}(t)\Big) = \hat{\alpha}_{i,k}T_{i}(t) + \hat{\beta}_{i,k} ,
\end{equation}
where $\hat{\alpha}_{i,k}$ and $\hat{\beta}_{i,k}$ are the clock frequency ratio
and the clock offset estimated based on the linear interpolation using four
timestamps, i.e.,
\begin{align}
  \label{eq:rsp_clock_ratio}
  \hat{\alpha}_{i,k} & = \dfrac{T1_{k}{-}T1_{k-1}}{T2_{k}{-}T2_{k-1}}, \\
  \label{eq:rsp_clock_offset}
  \hat{\beta}_{i,k} & = \dfrac{T1_{k-1}{\cdot}T2_{k}-T2_{k-1}{\cdot}T1_{k}}{T2_{k}-T2_{k-1}} .
\end{align}

\subsection{Impact of Limited Precision Floating-Point Arithmetic on
  High-Precision Time Synchronization}
\label{sec:precision_loss}
Even with the relatively simpler clock parameter estimation procedures, RSP
suffers from numerical errors caused by the limited precision of the low-cost
MCUs (i.e., 32-bit single-precision floating-point numbers) of the
resource-constrained sensor nodes: Specifically,
the proposers of RSP suggest to use SI (i.e., $T1-T1_{init}$ or $T2-T2_{init}$)
that is large enough to mitigate the impact of the numerical errors on the
estimation of clock parameters in \eqref{eq:rsp_clock_ratio} and
\eqref{eq:rsp_clock_offset}.

Likewise, the actual performance of EE-ASCFR on resource-constrained sensor
nodes turns out to be poorer than expected due to the precision loss resulting
from the use of single-precision floating-point numbers. In this regard, we have
proposed an improved version of the scheme along with its multi-hop extension
based on the reverse asymmetric framework and demonstrated satisfactory time
synchronization accuracy on a real WSN testbed \cite{Huan:19-1}.

It is worth mentioning that the use of 32-bit single-precision floating-point
format is common not only for the resource-constrained sensor node platforms
(e.g., TelosB \cite{telosb} and MicaZ \cite{MicaZ}) but also for the lightweight
Arduino boards \cite{Arduino}, which are frequently used as hardware platforms
for Internet of Things (IoT) prototyping as discussed in \cite{IoT:arduino}.

\section{Energy-Efficient Time Synchronization Tailored for Resource-Constrained
  Sensor Nodes}
\label{sec:energy-efficient_ts}
When the propagation delay is not significant (e.g., sub-microsecond delays for
WSNs with a communication range of \SI{300}{\m} or less), time synchronization
schemes based on the one-way message dissemination have a clear advantage over
those based on the two-way message exchange in terms of the number of message
transmissions at sensor nodes. Still, the schemes based on the one-way message
dissemination have issues in their implementation on resource-constrained sensor
nodes as discussed in Sec.~\ref{sec:oneway}. Here we introduce BATS---i.e., the
energy-efficient time synchronization scheme based on the reverse one-way
message dissemination---which addresses the implementation issues of the
one-way-message-dissemination-based time synchronization schemes on
resource-constrained sensor nodes.

\subsection{Impact of Precision Loss on the Performance of
  One-Way-Message-Dissemination-Based Time Synchronization Schemes}
\label{sec:issu-impl-one}
Taking RSP as an example, which was proposed to simplify the estimation of the
clock skew and offset in FTSP, we first analyze the impact of the precision loss
on the performance of the one-way-message-dissemination-based time
synchronization schemes.

The impact of the precision loss on the estimated reference clock of RSP in
\eqref{eq:rsp_logical_clock} can be analyzed in a similar way to that for
EE-ASCFR \cite{Huan:19-1}:
Let $\epsilon_{\alpha}$ and $\epsilon_{\beta}$ be the precision loss in the
estimation of the clock frequency ratio and the clock offset, i.e.,
\begin{align}
  \label{eq:precision_loss}
  \epsilon_{\alpha} & \triangleq \hat{\alpha}_{i,k} -
                      \hat{\alpha}_{i,k}^{\textnormal{LP}}, \\
  \epsilon_{\beta} & \triangleq \hat{\beta}_{i,k} - \hat{\beta}_{i,k}^{\textnormal{LP}},
\end{align}
where $\hat{\alpha}_{i,k}^{\textnormal{LP}}$ and
$\hat{\beta}_{i,k}^{\textnormal{LP}}$ are the value of the clock frequency ratio
and the clock offset from the practical implementation that are affected by the
limited-precision floating-point arithmetic in \eqref{eq:rsp_clock_ratio} and
\eqref{eq:rsp_clock_offset}. The computational error $\Psi(t)$ in the estimated
reference clock at the sensor node at time $t$, therefore, can be derived as
follows:
\begin{equation}
  \label{eq:computation_error}
  \begin{split}
    \Psi(t) & = \Big(\hat{\alpha}_{i,k}T_{i}(t) + \hat{\beta}_{i,k}\Big) -
    \Big(\hat{\alpha}_{i,k}^{\textnormal{LP}}T_{i}(t) +
    \hat{\beta}_{i,k}^{\textnormal{LP}}\Big), \\
    & = \epsilon_{\alpha}T_{i}(t) + \epsilon_{\beta} .
  \end{split}
\end{equation}
\eqref{eq:computation_error} shows that the computational error consists of two
components, the first of which is proportional to the current time of the sensor
node's hardware clock (i.e., $T_{i}(t)$).
%

According to the IEEE standard for floating-point arithmetic
\cite{IEEE:754-2008}, a non-zero floating-point number $x$ can be represented in
the binary format as follows:
\begin{equation}
  \label{eq:binary_float}
  x = \sigma\cdot\bar{b}\cdot2^e,
\end{equation}
where $\sigma$ is the \textit{sign} taking the value of ${+}1$ or ${-}1$,
$\bar{b}$ is the \textit{binary fraction} whose value is within the range of
$[1,2)$,
and $e$ is the \textit{integer exponent}.
Because the IEEE 32-bit single-precision floating-point format assigns 1 bit to
$\sigma$, 8 bits to $e$ and 23 bits to $\bar{b}$, the \textit{machine epsilon}
\cite{goldberg91:_what} becomes $2^{-23}$. If the rounding arithmetic is
chopping (i.e., rounding towards 0), the precision loss $\epsilon_{\alpha}$ is
within the range of $[{-}2^{-23},0]$, and the maximum absolute precision loss in
this case is $2^{-23}{\approx}\num{1.19e-7}$. This implies that the first
component of \eqref{eq:computation_error} alone could result in about
\SI{0.1}{\us} and \SI{1}{\us} computational errors for $T_{i}(t)$ of \SI{1}{\s}
and \SI{10}{\s}, respectively, in the worst case.

Note that the analysis of the computational error above is based on the
worst-case scenario for simplicity. As discussed in \cite{rsp}, however, in
reality the precision loss $\epsilon_{\alpha}$ itself is inversely proportional
to SI (i.e., the time difference between two consecutive beacon messages), which
could more or less relax the dependency of the computational error $\Psi(t)$ on
$T_{i}(t)$. In fact, setting an optimal value of SI is quite complicated because
the value of SI not only affects the computational error but also determines the
impact of the sensor node's hardware clock drift due to the changes in the
ambient temperature and the battery voltage.

\subsection{Beaconless Asymmetric Energy-Efficient Time Synchronization}
\label{sec:bats}
Now we introduce BATS and show how it can address the implementation issues on
resource-constrained sensor nodes discussed in Sec.~\ref{sec:issu-impl-one} and
further increase energy efficiency by saving the energy for transmissions and
receptions of the ``Beacon/Request'' messages at sensor nodes.

\begin{figure}[!tb]
  \centering \includegraphics[width=\linewidth]{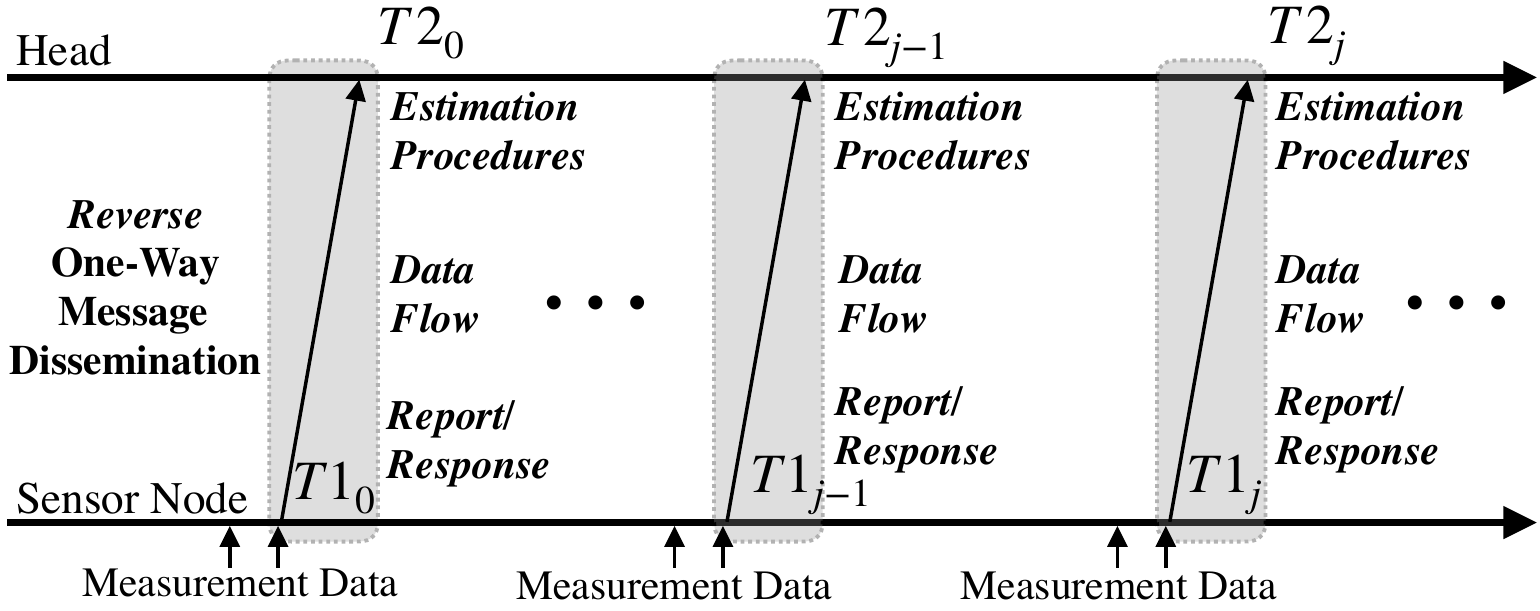}
  \caption{Asymmetric energy-efficient time synchronization based on reverse
    one-way message dissemination with optional bundling of measurement data.}
  \label{fig:reverse_oneway}
\end{figure}

\begin{figure}[!tb]
  \centering%
  \includegraphics[width=\linewidth]{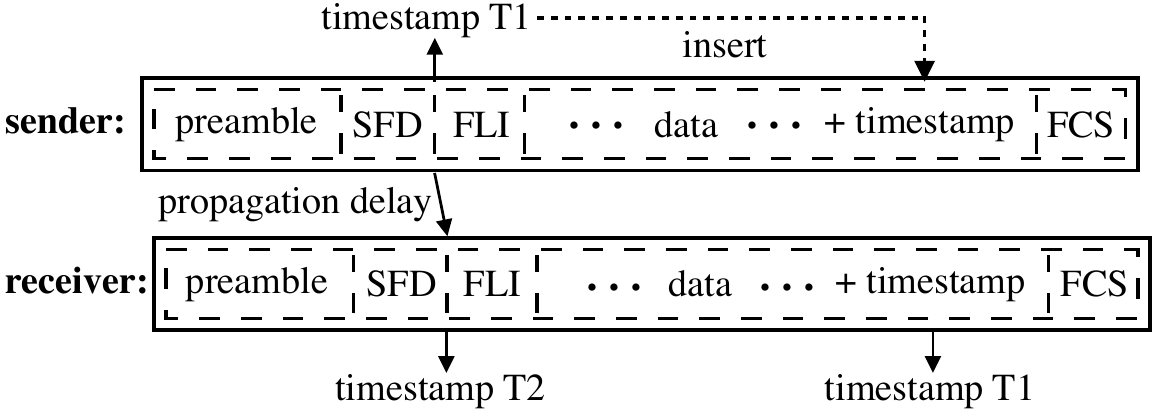}
  \caption{MAC-layer timestamping \cite{akhlaq13:_rtsp} adopted in BATS.}
  \label{fig:proposed_timestamping}
\end{figure}

As illustrated in Fig.~\ref{fig:reverse_oneway}, the proposed BATS scheme based
on the reverse one-way message dissemination does not rely on beacon messages
but utilizes only measurement data messages to carry the reversed
synchronization timestamp $T1$, which results in the movement of all time
synchronization procedures to the head except timestamping of $T1$. As shown in
Fig.~\ref{fig:proposed_timestamping}, the timestamping of $T1$ and $T2$ is
triggered by an interrupt generated by the radio chip immediately after the
Start Frame Delimiter (SFD) byte has been sent and received by the MAC layer of
the sender and the receiver, respectively, which is similar to that of the
Recursive Time Synchronization Protocol (RTSP) \cite{akhlaq13:_rtsp}. This
timestamping approach, which is based on a pair of timestamps generated during
the transmission and reception of one message, is simpler and faster than that
of FTSP \cite{akhlaq13:_rtsp} which reduces the jitter of interrupt handling
through recording, normalizing and averaging multiple timestamps at both the
sender and the receiver.

In the single-hop scenario shown in Fig.~\ref{fig:reverse_oneway}, the head's
direct offspring nodes employ their measurement data messages to transmit their
timestamps $T1_{j}$ during the $j$-th synchronization to the head for maintaining
the time synchronization. Using the abundant computing and power resources
including the 64-bit floating-point precision at the head, the numerical
computational errors mentioned in Section.~\ref{sec:related_work} could be
avoided in the proposed scheme.

Instead of using the relatively simpler clock parameter
estimation procedures of \cite{Kim:17-1} and \cite{rsp}, we employ the original 
linear regression like \cite{maroti04} and \cite{elson02:_fine} to model the 
relationship of the timestamps of the hardware clocks of the sensor node and the 
head, however, without the limitation on the number of past samples, e.g., FTSP 
uses only the past 8 samples. The following linear least squares method is employed 
in BATS to model the linear relationship between the hardware clocks of the sensor 
node $i$ and the head: For $i{\in}\left[0,1,\ldots,N{-}1\right]$
\begin{equation}
  \label{eq:normal_equation}
  \Phi_{i}(j) = (T2_{i}(j)^{T}T2_{i}(j))^{-1}T2_{i}(j)^{T}T1_{i}(j),
\end{equation}
where
\[
\begin{split}
	\Phi_{i}(j)=[1+\hat{\epsilon}_{i}(j), \hat{\theta}_{i}(j)],\\
	T1_{i}(j)=[T1_{i}(j-m+1), ... , T1_{i}(j)],\\
	T2_{i}(j)=[T2_{i}(j-m+1), ... , T2_{i}(j)].
\end{split}
\]
where $\Phi_{i}(j)$ is the parameter set of clock skew and offset, and $j$ 
denotes the $j$-synchronization.

As we will see in Section~\ref{sec:exp_results}, employing the complex linear
regression solution with no limitation on the sample size for solving the linear
equation of the time synchronization, could improve the performance of time
synchronization compared with the conventional simpler solutions described in
Section~\ref{sec:related_work}. Furthermore, applying relatively more complex
solution with higher computational complexity is the natural advantage of our
reverse asymmetric framework as we operate all the estimation procedures in the
head with abundant computing capability instead of the resource-constrained
sensor node.

As exhibited in Fig.~\ref{fig:reverse_oneway}, the direction of the conventional
one-way message dissemination is reversed, where the estimation of the hardware
clocks between sensor node and the head is also reversed compared to the
conventional one. On the one hand, when timestamps along with measurement data
from sensor nodes are gathered in the head, the corresponding hardware clock
timestamps of the sensor nodes should be translated based on the head's
reference clock to estimate the exact time for the measurement events. On the
other hand, optionally, when the head issues commands to notify sensor nodes to perform
collaborative operations (e.g., sleep and wake-up for energy-efficient MAC
protocols and application of coherent sampling), the timestamps carried in the
commands have to be translated to the target sensor nodes' hardware clock
time. Using the estimated frequency ratio $1+\epsilon_{i}$ and offset
$\theta_{i}$ between a sensor node $i$ and the head, in single-hop scenario, the
translations between their hardware clock time are represented as:
\begin{equation}
  \label{eq:timestamp_translation_singlehop}
  t = \frac{T_{i}(t) - \theta_{i}}{1+\epsilon_{i}},~~ T_{i}(t) = (1+\epsilon_{i})t + \theta_{i},
\end{equation}
where the translation from sensor node's clock time $T_{i}(t)$ to head's clock
time $t$ involved the floating-point division, which is similar to the clock
time translation in EE-ASCFR. As mentioned in \cite{Dja14:_imp}, such
calculation involved with floating-point division and microsecond-level
timestamp should be carefully handled in the resource-constrained sensor nodes,
but BATS is not bothered by such level of computational complexity.

\section{Extension of BATS to Multi-Hop WSNs}
\label{sec:multi-hop extension}
Independent of the network topologies which are typically established through
routing protocols (e.g., Collection Tree Protocol (CTP) \cite{ctp}), the
proposed BATS scheme in principle could be extended to any existing multi-hop
routing protocols with the addition of the required timestamps in the
application messages. There are several issues to consider, however, in its
extension to the multi-hop scenario.

In the multi-hop scenario, gateway sensor nodes located in intermediate layers
process not only their own data as regular sensor nodes but also the data from
their offspring sensor nodes. The presence of gateway nodes makes it
complicated for the head to directly handle all timestamps required for time
synchronization in multi-hop WSNs.

\subsection{Multi-Hop Extension of The Time Synchronization Based On Reverse
  One-Way Message Dissemination}
\label{sec:extension_reverse_one_way}
\begin{figure}[!tb]
  \centering%
  \includegraphics[width=\linewidth]{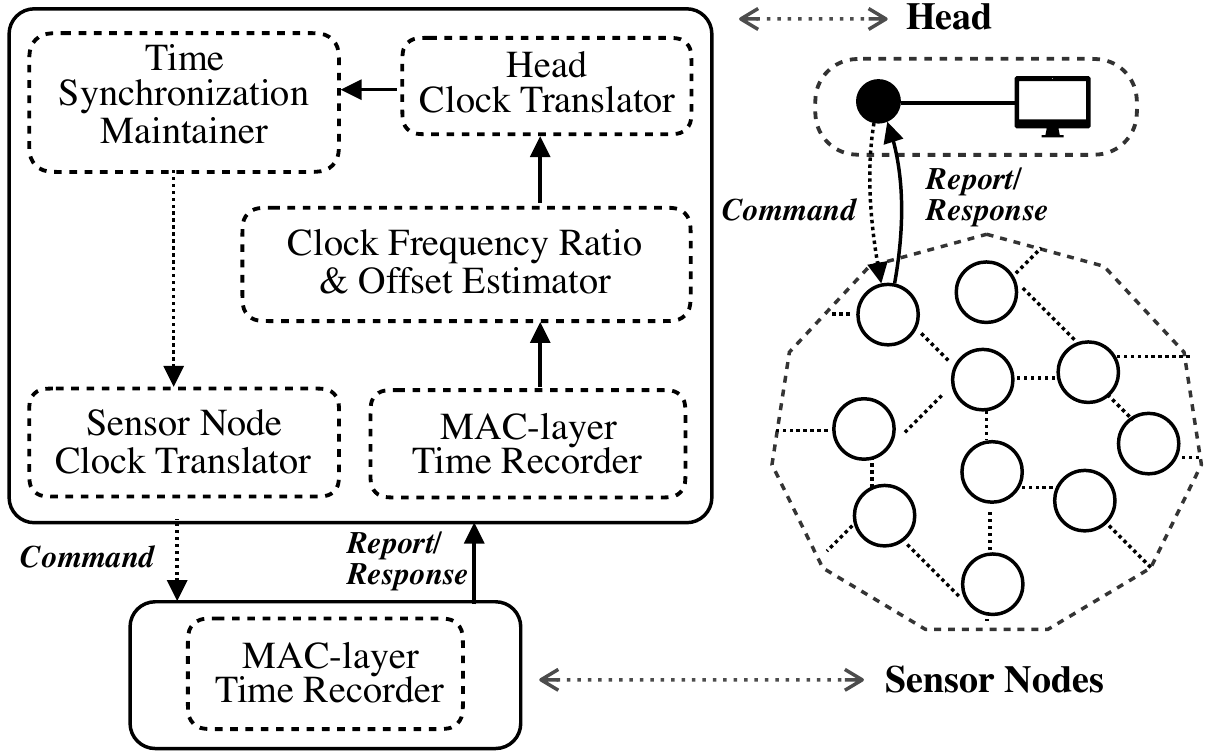}
  \caption{A system architecture of the proposed time synchronization scheme.}
  \label{fig:bats_architecture}
\end{figure}

As described in \cite{Kim:17-1}, two possible approaches for the multi-hop
extension of WSN time synchronization schemes based on the reverse two-way
message exchange (including EE-ASCFR) are those of \textit{time-relaying} and
\textit{time-translation} at the gateway nodes. Of the two approaches, the
time-relaying one could introduce more random delays as the messages from the
sensor nodes are being transmitted to the head through multiple gateway nodes
unlike the time-translation one. Moreover, the random queueing delays of the
synchronization messages are also not properly compensated for. To maximize the
advantage of MAC-layer timestamping and avoid random queueing delays cumulated
through multiple gateway nodes, therefore, we use \textit{per-hop time
  synchronization} for the multi-hop extension of BATS, which is similar to the
aforementioned time-translation approach in which the hardware clock time of a
sensor node goes through layer-by-layer translation in order to estimate its
time with respect to the reference clock of the head. The fundamental difference
of the per-hop time synchronization compared to the time-translation approach,
however, is that all layer-by-layer translations are again moved from the
gateway nodes to the head to relieve the burden of the gateway nodes in the
original time-translation approach. The system architecture is illustrated in
Fig.~\ref{fig:bats_architecture}, where the ``command'' is optional and the 
``sensor nodes'' refer to both gateway and leaf sensor nodes.
\begin{figure*}[t]
  \setlength{\myvspace}{0.05cm}%
  \begin{center}
  	\includegraphics[width=.65\linewidth]{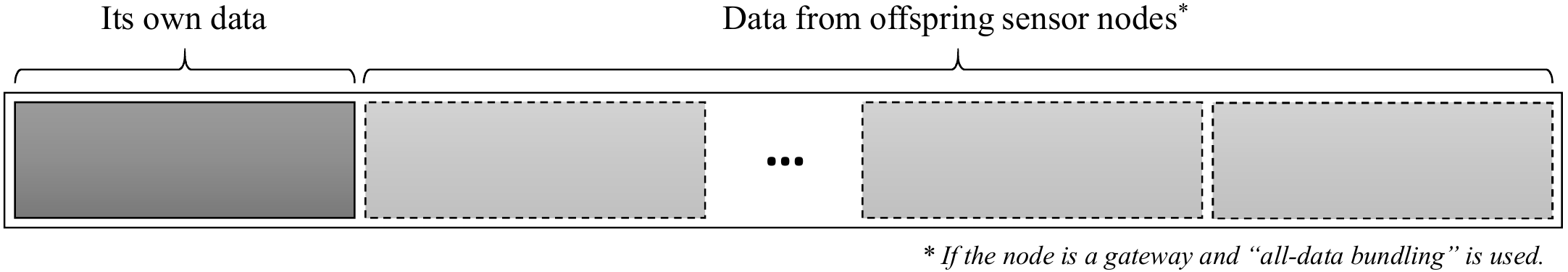} \\
    \vspace{\myvspace}%
    {\scriptsize (a)}
  \end{center}
  \begin{center}
  	\includegraphics[width=.7\linewidth]{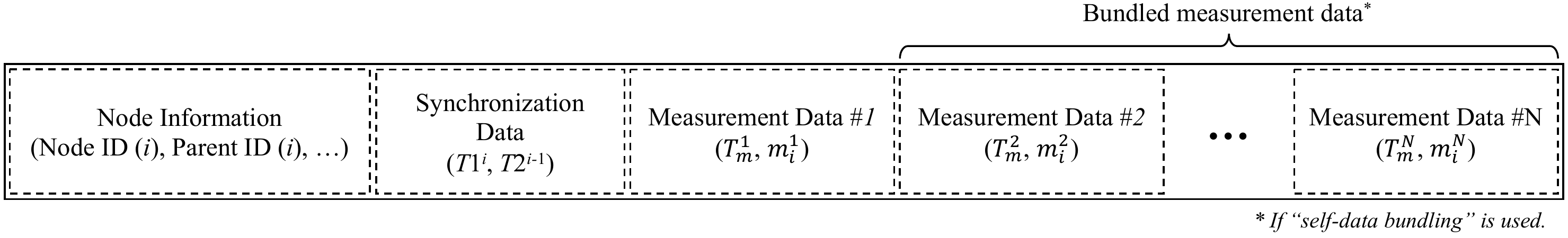} \\
    \vspace{\myvspace}%
    {\scriptsize (b)}
  \end{center}
  \caption{Payload and data structure of a message generated at sensor node $i$:
    (a) Payload with optional ``all-data bundling'' and (b) data structure with
    optional ``self-data bundling''.}
  \label{fig:payload}
\end{figure*}

We first extend the hardware clock time translation equation of
\eqref{eq:timestamp_translation_singlehop} for the single-hop case to the
multi-hop case where we recursively translate the hardware clock time. Consider
the hardware clock time of a sensor node located at layer $j$ of the sensor
network with $n$ layers: For $j{\in}\left[0,1,\ldots,n\right]$,
\begin{equation}
  \label{eq:timestamp_translation_multihop_sensor2head}
  T^{j}_{i}(t)= T^{n-k}_{i}(t)=
  \begin{cases}
    t,& \text{$k = n$}\\
    \frac{T^{n-k+1}_{i}(t) - \theta^{n-k+1}_{i}}{1+\epsilon^{n-k+1}_{i}},&
    \text{$0 < k < n$}
  \end{cases}
\end{equation}
where $T^{j}_{i}(t)$ is the hardware clock time of the head when $k$ is equal
to $n$.  And the equation when $0 < k < n$ represents the layer-by-layer
translation using the clock frequency ratio $1+\epsilon^{n-k+1}_{i}$ and offset
$\theta^{n-k+1}_{i}$ of each layer.

To translate the hardware clock time of a sensor node at layer $j$ to that based
on the hardware clock time of the head (i.e., $t$), we go through the following
recursive procedure that is to run at the head:
\begin{equation}
  \label{eq:timestamp_translation_multihop_head2sensor}
  T^{j}_{i}(t)=
  \begin{cases}
    (1+\epsilon^{j}_{i})t + \theta^{j}_{i},& \text{$j$ = 1}\\
    (1+\epsilon^{j}_{i})T^{j-1}_{i}(t) + \theta^{j}_{i},& \text{1 < $j$ $\leq$
      $n$}
  \end{cases}
\end{equation}
where the clock frequency ratio $1+\epsilon^{j}_{i}$ and clock offset
$\theta^{j}_{i}$ of each layer are all known at the head.

Note that, in the proposed per-hop time synchronization, timestamps required for
the synchronization of a sensor node based on the reverse one-way message
dissemination are recorded at the sensor node and its gateway node (working as a
reference) as shown in Fig.~\ref{fig:reverse_oneway}. Unlike the time
translation approach described in \cite{Kim:17-1}, however, the time translation
operation is moved from the gateway to the head, and the gateway node just sends
the pair of timestamps obtained from the reverse one-way message dissemination
(i.e., $T1_{i}$ and $T2_{i-1}$) to the head. Based on these pairs of timestamps
from all the gateway nodes, the head can establish the relationships between
sensor nodes and their gateway nodes and eventually translate sensor nodes
hardware clock times to those based on the reference clock at the head
recursively based on \eqref{eq:timestamp_translation_multihop_sensor2head} and
\eqref{eq:timestamp_translation_multihop_head2sensor}.

\begin{figure}[!tb]
  \centering%
  \includegraphics[width=\linewidth]{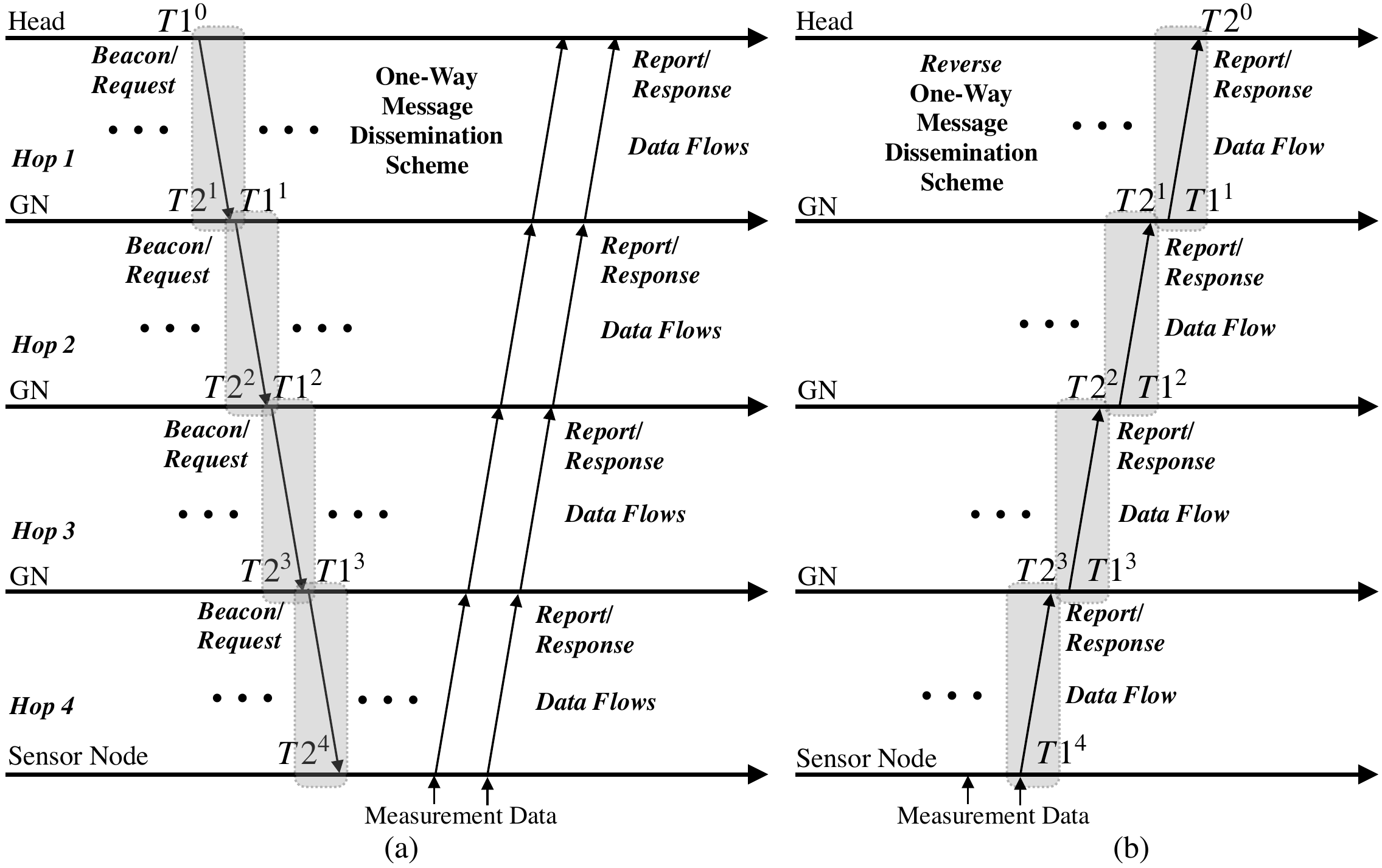}
  \caption{Multi-hop extension of (a) a conventional (e.g., FTSP) and (b) a
    reverse asymmetric time synchronization scheme based on the one-way message
    dissemination with all-data bundling procedure.}
  \label{fig:multihop_one_way}
\end{figure}

Using the aforementioned per-hop time synchronization, BATS could be extended to
cover the multi-hop scenario as demonstrated in
Fig.~\ref{fig:multihop_one_way}~(b), which is different from the conventional
multi-hop scheme based on one-way message dissemination shown in
Fig.~\ref{fig:multihop_one_way}~(a) where the synchronization timestamps are
carried by standalone beacon messages and the measurement data are transported
by the standalone measurement messages. The proposed BATS does not rely on
broadcasting beacon messages and embeds the synchronization timestamps into the
measurement messages as in \cite{Kim:17-1}.
In case of the conventional schemes, there are two separate flows of messages
with different directions (i.e., one for synchronization and the other for
measurement), we cannot embed synchronization timestamps into the measurement
messages as in the extended BATS.

\subsection{Comparison to Other Multi-Hop WSN Time Synchronization Schemes}
\label{sec:comparison}
Having discussed the multi-hop extension of BATS, which does not rely on beacon
messages but exploits the bundling of synchronization timestamps with the
measurements, we compare the extended BATS to other multi-hop WSN time
synchronization schemes.

Since the synchronization timestamps are bundled with measurement data in the
same message, consequently the major cost of using the proposed scheme is the
payload occupation of the synchronization timestamp---i.e., $T1$ and $T2$---on
the measurement message. As exhibited in Fig.~\ref{fig:multihop_one_way}~(b),
for hop $j$, $2$ timestamps---i.e., $T1^{j}$ and $T2^{j-1}$---are required for
the head to achieve network-wide time synchronization. Therefore, for a flat
$n$-hop network, $2n{-}1$ timestamps\footnote{Because $T2^{0}$ is generated at
  the head, there is no need of transmission.}  are transmitted through the
measurement messages to the head. Note that, those timestamps ($T1$ and $T2$)
are also needed in the conventional scheme based on one-way message
dissemination such as FTSP, however, the cost---i.e., the transmissions of the
standalone synchronization messages---is much higher. Even considering that the
timestamp $T2$ could be kept locally in the sensor node to perform the
synchronization procedures, for a flat $n$-hop network, $n$ synchronization
messages which carry the timestamp $T1$ have to be broadcasted for achieving
overall time synchronization.
%
Using the self-data and all-data bundling procedure illustrated in Fig.~\ref{fig:payload}, 
for a flat $n$-hop network with generating $m$ measurements each hop, the numbers 
of message receptions and transmissions for conventional scheme ($M_{conv}$) and 
proposed scheme ($M_{prop}$) with bundling procedures could be quantified as follows:
\begin{equation}
  \label{eq:message_rx_tx_conv}
  M_{conv}= 2(n-1)+1+m\sum_{i=1}^{n}\Big(2(i-1)+1\Big)
\end{equation}
\begin{equation}
  \label{eq:message_rx_tx_prop}
  M_{prop}=
  \begin{cases}
  	\sum\limits_{i=1}^n(2(i-1)+1),& \text{self-data bundling}\\
  	2(n-1)+1,& \text{all-data bundling}
  \end{cases}
\end{equation}
where $M_{conv}$ contains not only the synchronization message but also 
the measurement message. On the contrary, $M_{prop}$ just involves the 
measurement message with synchronization timestamps inside. A flat $4$-hop 
network with generating $2$ measurements each hop is taken as an example with 
applying \eqref{eq:message_rx_tx_conv} and \eqref{eq:message_rx_tx_prop}, the 
numbers of message transmissions and receptions for conventional scheme and 
proposed scheme with self-data and all-data bundling procedures are $39$, 
$16$ and $7$. The proposed scheme with all-data bundling procedure conserves 
more than $80\%$ message transmissions and receptions compare to the conventional 
scheme.

Compared to the multi-hop extension of the time synchronization scheme based on
reverse two-way message exchange proposed in \cite{Huan:19-1}, the multi-hop
extension of BATS based on one-way message dissemination can greatly lower
communication overheads by reducing the number of synchronization messages
required for network-wide synchronization, which is a significant advantage
when used for large-scale WSNs. If the end-to-end communication range of a
multi-hop WSN is over a kilometer, however, the propagation delay cannot be
ignored any longer, and the time synchronization schemes based on reverse
two-way messages exchange would be a better solution in such a case.

%


\section{Experimental Results}
\label{sec:exp_results}
The proposed BATS scheme is implemented on a real testbed for a flat $6$-hop WSN
consisting of one head and six sensor nodes, all of which are based on TelosB
motes running TinyOS \cite{tinyos}. Note that the timer resolution of TelosB
motes running TinyOS is \SI{1}{\us} since the resolutions of its hardware clock
running on a 32-kHz crystal oscillator and software timer are \SI{30.5}{\us} and
\SI{1}{\us}, respectively. The accuracy of time synchronization schemes tested
on the testbed, therefore, is limited to a microsecond level.

\subsection{Energy Efficiency}
\label{sec:energy_efficiency}

\subsubsection{The Number of Message Transmissions/Receptions}
\label{sec:message_comparison}
\begin{table}[!tb]
  \centering
  \begin{threeparttable}
    \caption{The Numbers of Message Transmissions and Receptions at Sensor Node
      for Different SIs During The Period of \SI{3600}{\s}}
    \label{tab:message_comparison}
    \setlength{\tabcolsep}{4mm}%
    \centering
    \begin{tabular}{|c|l||r|r|r|}
      \hline
      \multicolumn{2}{|c||}{Type of Synchronization Scheme}
      & N$_{\rm TX}$\tnote{1}
      & N$_{\rm RX}$ \\ \hline\hline
      \multirow{3}{*}{Conventional Two-way}
      & SI$\;=100~\mbox{s}$  & 136 & 36\\ \cline{2-4}
      & SI$\;=10~\mbox{s}$  & 460 & 360\\ \cline{2-4}
      & SI$\;=1~\mbox{s}$  & 3700 & 3600\\ \hline\hline
      \multirow{3}{*}{Conventional One-way}
      & SI$\;=100~\mbox{s}$  & 100 & 36\\ \cline{2-4}
      & SI$\;=10~\mbox{s}$  & 100 & 360\\ \cline{2-4}
      & SI$\;=1~\mbox{s}$  & 100 & 3600\\ \hline\hline
      \multirow{3}{*}{Reverse Two-way}
      & SI$\;=100~\mbox{s}$  & 100 & 36\\ \cline{2-4}
      & SI$\;=10~\mbox{s}$  & 100 & 360\\ \cline{2-4}
      & SI$\;=1~\mbox{s}$  & 100 & 3600\\ \hline\hline
      \multirow{3}{*}{Reverse One-way}
      & SI$\;=100~\mbox{s}$  & 100 & 0\\ \cline{2-4}
      & SI$\;=10~\mbox{s}$  & 100 & 0\\ \cline{2-4}
      & SI$\;=1~\mbox{s}$  & 100 & 0\\ \hline
    \end{tabular}
    \begin{tablenotes}
    \item[1] Both synchronization and measurement messages are counted.
    \end{tablenotes}
  \end{threeparttable}
\end{table}

%
%

First, we count the numbers of message transmissions and receptions of the
conventional and reverse asymmetric time synchronization schemes based on both
one-way message dissemination and two-way message exchange to indirectly compare
their energy consumptions. A single-hop scenario is considered, where the head
and one sensor node are directly connected to each other. We assume that there
are $100$ measurements in total at the sensor node over the period of
\SI{3600}{\s}, which are reported to the head through measurement messages
without measurement bundling, and the SI is set to \SI{1}{\s}. The
resulting number of messages---i.e., synchronization and measurement
messages---transmitted (N$_{\rm TX}$) and received (N$_{\rm RX}$) by the sensor
node are summarized in Table.~\ref{tab:message_comparison}.

From the comparison, we observe that the reverse one-way scheme could save the
energy consumed by both transmissions and receptions of the synchronization
messages; this is because the reverse one-way scheme is free from transmissions
and receptions of beacon messages and synchronization timestamps are embedded
into measurement messages.

In the reverse two-way scheme, the ``Request'' synchronization message including
timestamp $T1$ can be embedded into beacon messages, but still sensor nodes
consume energy to receive the beacon messages. On the other hand, the
conventional two-way scheme requires the most message transmissions due to the
sending the ``Request'' messages from the sensor node to the head. In addition,
the conventional one-way scheme requires the same number of synchronization
message receptions as in the reverse two-way scheme.

\subsubsection{Direct Measurement of Energy Consumption}
\label{sec:energy_consumption}
\begin{figure}[!tb]
  \centering%
  \includegraphics[width=\linewidth]{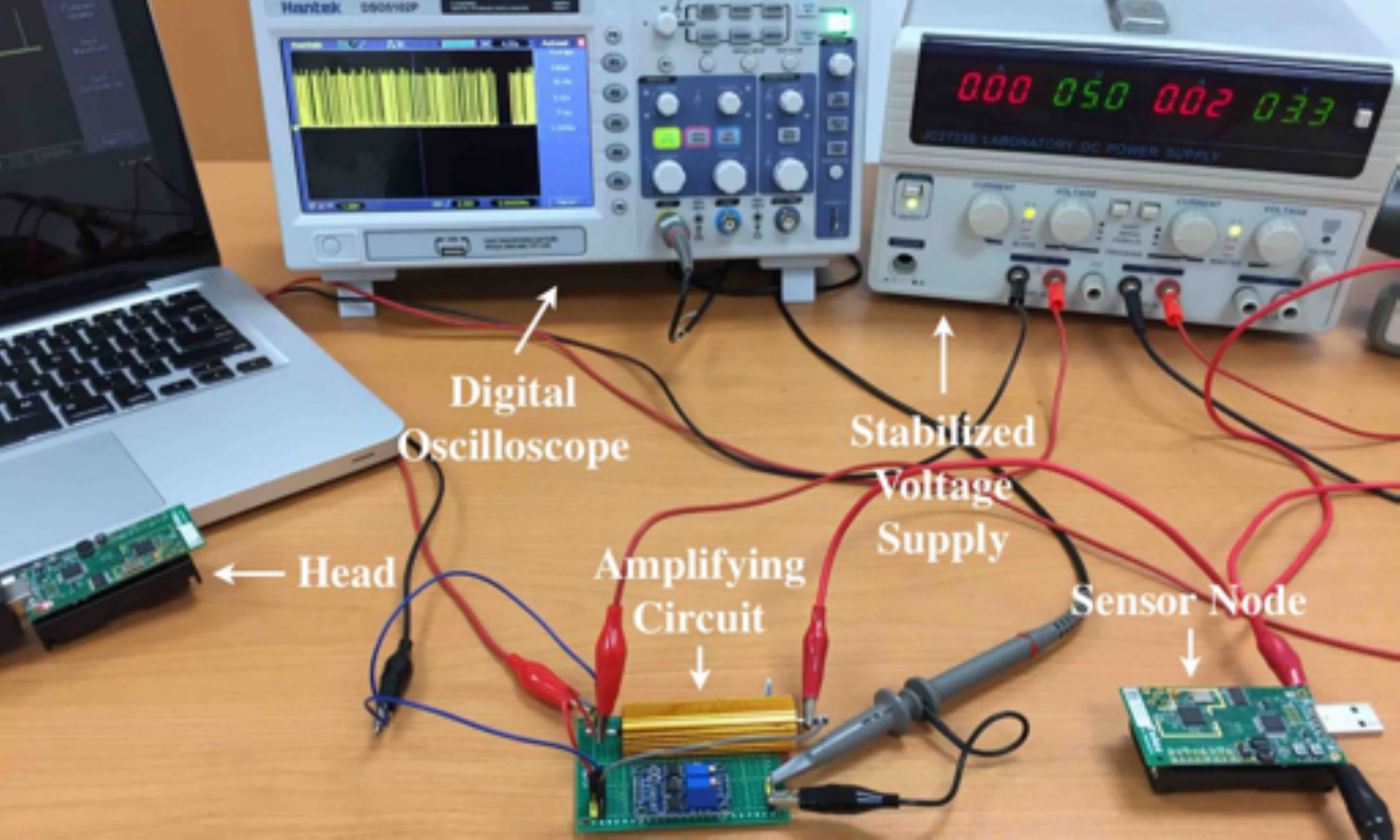}
  \caption{Experiment setup for the measurement of the energy consumption on a
    WSN sensor node.}
  \label{fig:experiment_setup}
\end{figure}

\begin{figure}[!tb]
  \centering%
  \includegraphics[width=\linewidth]{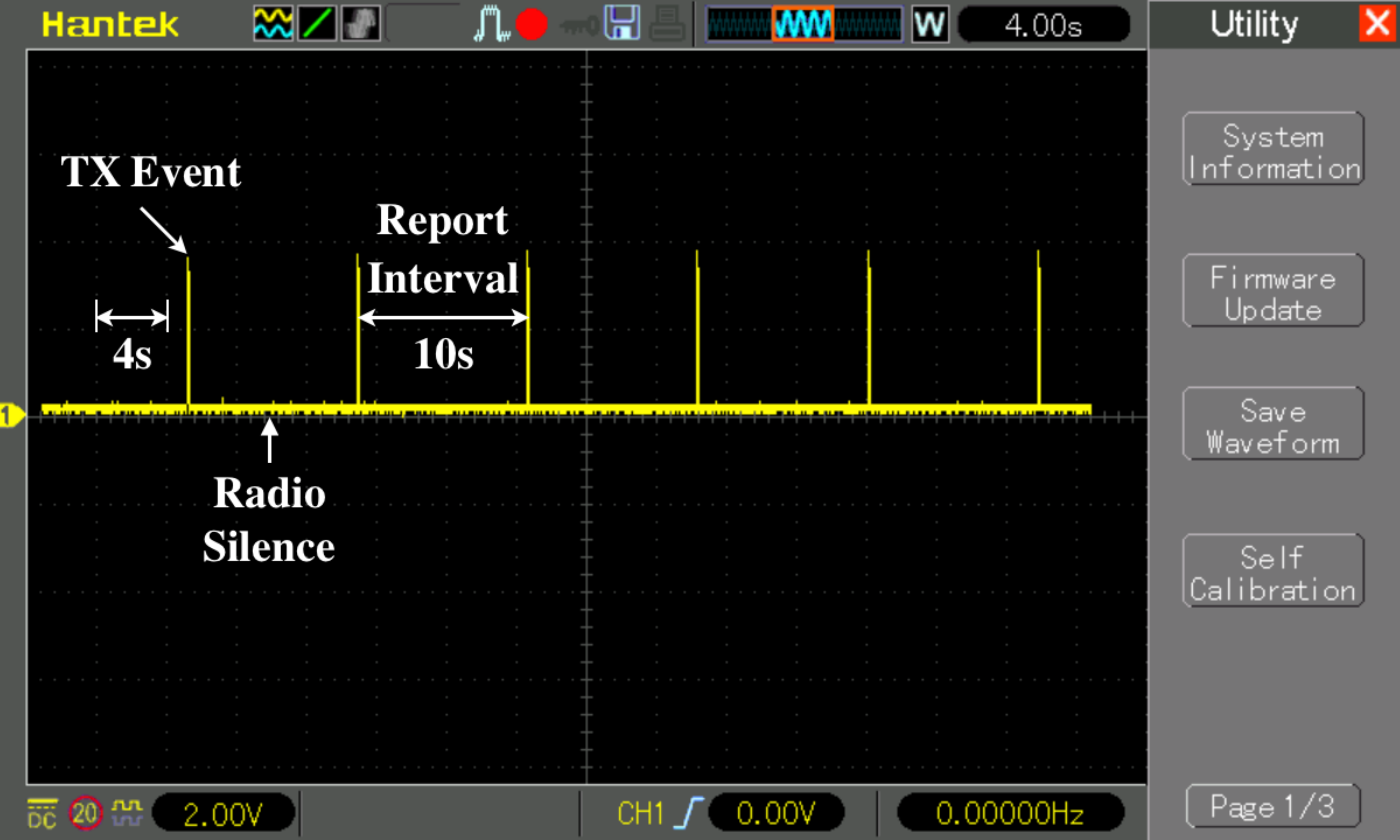}
  \caption{Measuring and logging the energy consumption using DSO.}
  \label{fig:dso}
\end{figure}

To measure the actual power consumption of a time synchronization scheme on the
resource-constrained sensor nodes, we employ a stabilized voltage supply and a
digital storage oscilloscope (DSO) to power the sensor node which synchronizes
with the head and log the actual power consumption as shown in
Fig.~\ref{fig:experiment_setup}. In the experiment setup, one \SI{1}{\ohm}
resistor is connected in series between the \SI{3.3}{\volt} power supply and the
sensor node. The two pins of the resistor are connected to the inputs of the
amplifying circuit\footnote{The amplifying circuit is based on the analog
  devices designer guidebook \cite[chapter~4]{AD_BOOK}.} which used to enlarge
the \si{\mA}-level current input value with 150 times. The outputs of the
amplifying circuit are connected to the DSO for logging the power
consumption. The power consumption logged in the DSO is illustrated in
Fig.\ref{fig:dso}. Because the voltage is stabilized in the measurement
experiment which is $3.3V$ constantly, so the actual power consumption could be
compared through comparing the logged current value sets.

\begin{figure}[!tb]
  \centering%
  \includegraphics[width=\linewidth]{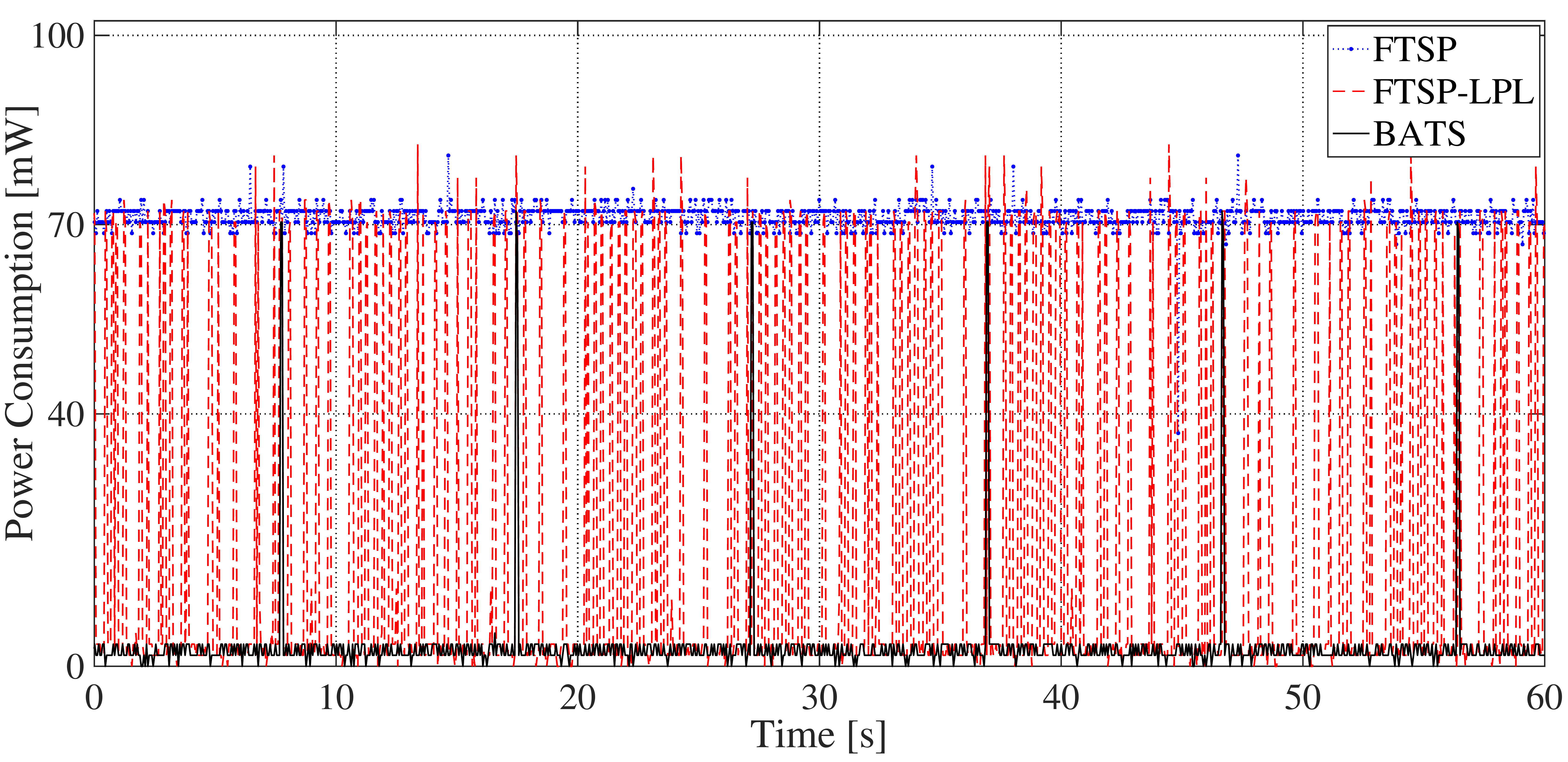}
  \caption{Power consumptions of different time synchronization schemes
    over \SI{60}{\s}.}
  \label{fig:current_time}
\end{figure}

\begin{figure}[!tb]
  \centering \includegraphics[width=\linewidth]{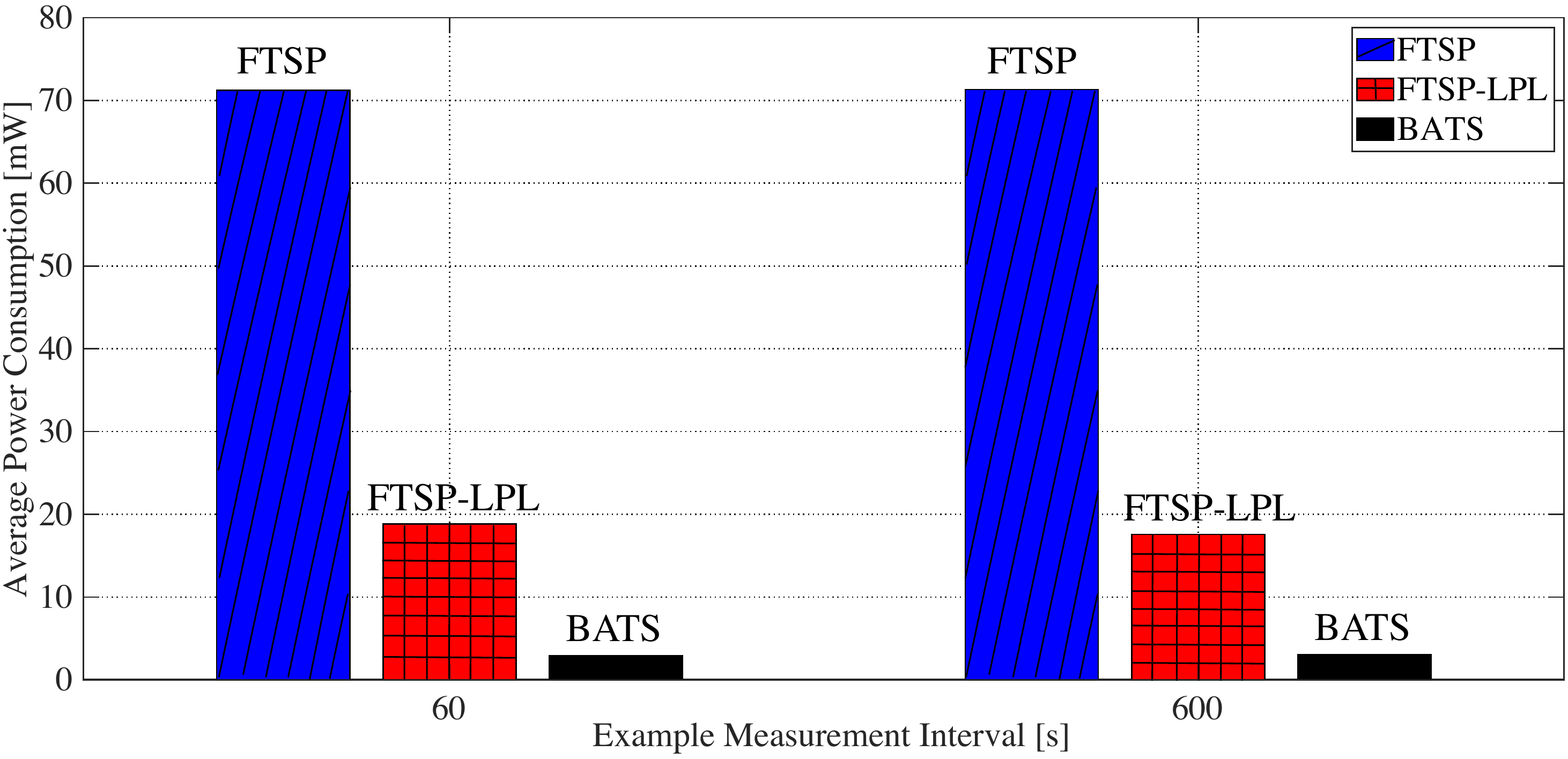}
  \caption{Average power consumptions of time synchronization schemes over the
    measurement interval of \SI{60}{\s} and \SI{600}{\s}.}
  \label{fig:average_power}
\end{figure}

During the experiments, the measurement is generated every \SI{2}{\s} and the 
bundling procedure with bundling $5$ measurements is employed in both 
FTSP and BATS for fair comparison. FTSP broadcasts and receives the 
synchronization beacon messages per second (i.e., default setting) and reports the 
bundling messages every \SI{10}{\s}, BATS reports the bundling messages with the 
measurements and synchronization data every \SI{10}{\s} without broadcasting or 
receiving---i.e., radio listening deactivated---the synchronization beacons. 
The power and energy consumptions are computed as follows:
\begin{align}
  P(t) & = V(t) \times I(t), \label{eq:power_consumption}\\
  E & = \int_{t_{s}}^{t_{e}}P(t)dt, \label{eq:energy_consumption}
\end{align}
where $P(t)$, $V(t)$ and $I(t)$ are the instant power, voltage and current for
the sensor node at time $t$, respectively, and
$E$ is the energy consumed by the sensor node over the time period of
$\mathopen[t_{s},t_{e}\mathclose]$.
Because the voltage $V(t)$ is fixed to \SI{3.3}{\volt} during the experiments,
we only measure the current $I(t)$ over the two different time intervals of
\SI{60}{\s} and \SI{600}{\s} (i.e., $t_{e}{-}t_{s}$ in
\eqref{eq:energy_consumption}) to obtain power and energy consumptions;
Fig.~\ref{fig:current_time} shows the power consumption for different
time synchronization schemes over the period of \SI{60}{\s}. Since the
broadcasting of beacon messages and the reporting of the (bundled) measurement
message are periodic, so the two aforementioned example measurement intervals
could represent the long-term experiments. As illustrated in
Fig.~\ref{fig:average_power}, the average power consumption for BATS is the
lowest, while the original FTSP---i.e., without using low-power mode---consumes
the most power. Specifically, the FTSP employing low-power mode consumes less
than one-third power of the original FTSP but more than BATS, which consumes
less than $5\%$ and $16\%$ power of the original and the low-power FTSP,
respectively.

\subsection{Time Synchronization Accuracy}
\label{sec:synchronization_performance}

\subsubsection{Single-Hop Scenario}
\label{sec:synchronization_performance_single}
We first evaluate the time synchronization accuracy of the proposed BATS scheme
in comparison to that of FTSP with a single-hop scenario. For the experiments,
we assume that the sensor node periodically sends measurement data to the head
via measurement messages, where 5 measurements are bundled in each measurement 
message for both FTSP and BATS together with synchronization data in case of the 
latter. We run experiments for \SI{3600}{\s}.

Note that BATS uses the measurement messages to carry both measurement and
synchronization data, while conventional one-way time synchronization schemes
like FTSP rely on beacon messages for synchronization data and use measurement
messages only for measurement data. For a fair comparison between BATS and
FTSP, therefore, we define SI as the interval between two consecutive
measurement messages carrying synchronization data for BATS and the interval
between two consecutive beacon messages for FTSP, respectively, and use the same
SI value for both schemes for each experiment.


%

\begin{table*}[t]
  \centering
  \begin{threeparttable}
    \caption{MAE and MSE of Measurement Time Estimation of FTSP and BATS for the
      Single-Hop Scenario}
    \label{tab:evaluation_results_sis}
    \setlength{\tabcolsep}{7mm} \centering
    \begin{tabular}{|c|l||r|r|r|r|r|}
      \hline
      \multicolumn{2}{|c||}{Synchronization Scheme}
      & \multicolumn{1}{c|}{MAE\tnote{1} [s]}
      & \multicolumn{1}{c|}{MSE\tnote{1}}
      & \multicolumn{1}{c|}{N$_{\rm TX}$}
      & \multicolumn{1}{c|}{N$_{\rm RX}$} \\ \hline\hline
      \multirow{3}{*}{FTSP\tnote{2}}
      & SI$\;=100~\mbox{s}$  & 0.2892E-03 & 0.3614E-06 & 36 & 36\\ \cline{2-6}
      & SI$\;=10~\mbox{s}$  & 0.3164E-03 & 0.3983E-06 & 360 & 360\\ \cline{2-6}
      & SI$\;=1~\mbox{s}$  & 0.3173E-03 & 0.4038E-06 & 3600 & 3600\\ \hline\hline
	  \multirow{3}{*}{BATS with the method 1}
      & SI$\;=100~\mbox{s}$  & 2.4837E-05 & 2.1194E-09 & 36 & 0\\ \cline{2-6}
      & SI$\;=10~\mbox{s}$  & 3.2770E-06 & 1.7492E-11 & 360 & 0\\ \cline{2-6}
      & SI$\;=1~\mbox{s}$  & 2.4903E-06 & 1.0120E-11 & 3600 & 0\\ \hline\hline
      \multirow{3}{*}{BATS with the method 2}
      & SI$\;=100~\mbox{s}$  & 8.1524E-06 & 1.5805E-10 & 36 & 0\\ \cline{2-6}
      & SI$\;=10~\mbox{s}$  & 2.1016E-06 & 7.3933E-12 & 360 & 0\\ \cline{2-6}
      & SI$\;=1~\mbox{s}$  & 1.8299E-06 & 5.4018E-12 & 3600 & 0\\ \hline
    \end{tabular}
    \begin{tablenotes}
    \item[1] Based on the measurement time estimation obtained from
      \SI{3600}{\s} such that the actual performance in real deployment is
      represented.
    \item[2] The standard FTSP implementation provided in TinyOS library offers
      limited millisecond-level time synchronization
      \cite{Temperature-Compensated}.
    \end{tablenotes}
  \end{threeparttable}
\end{table*}

Before carrying out a comparative analysis with FTSP, we need to decide the
values of BATS system parameters affecting time synchronization
performance. During the preliminary experiments, we found that the time
synchronization accuracy of BATS is affected by the sample size for the linear
regression for clock frequency ratio estimation. Note that, for a reasonable
comparison, to find the best value of the sample sizes for different SIs, we
implemented the offline re-running program in which the same raw data trace
produced in the online experiment could be reused to run different offline
experiments with different sample sizes. With this program, the possible sample
sizes (e.g., 2, \dots , 10, \dots, 100, \dots, all) for different values of SI
are evaluated. In particular, the performance of the experiment based on all
samples does not outperform the one using the most recent timestamps with
certain sample size value. This may result from, as time goes on, due to the
aggravating clock drift, the very past sample data---i.e., very past
timestamps---do not have positive contributions for the estimations of the
timestamps in the most current synchronization interval. Instead, the most
recent samples could provide relatively better contributions for the estimation
of the most recent timestamps.

\begin{figure}[!tb]
  \begin{center}
    \includegraphics[width=\linewidth]{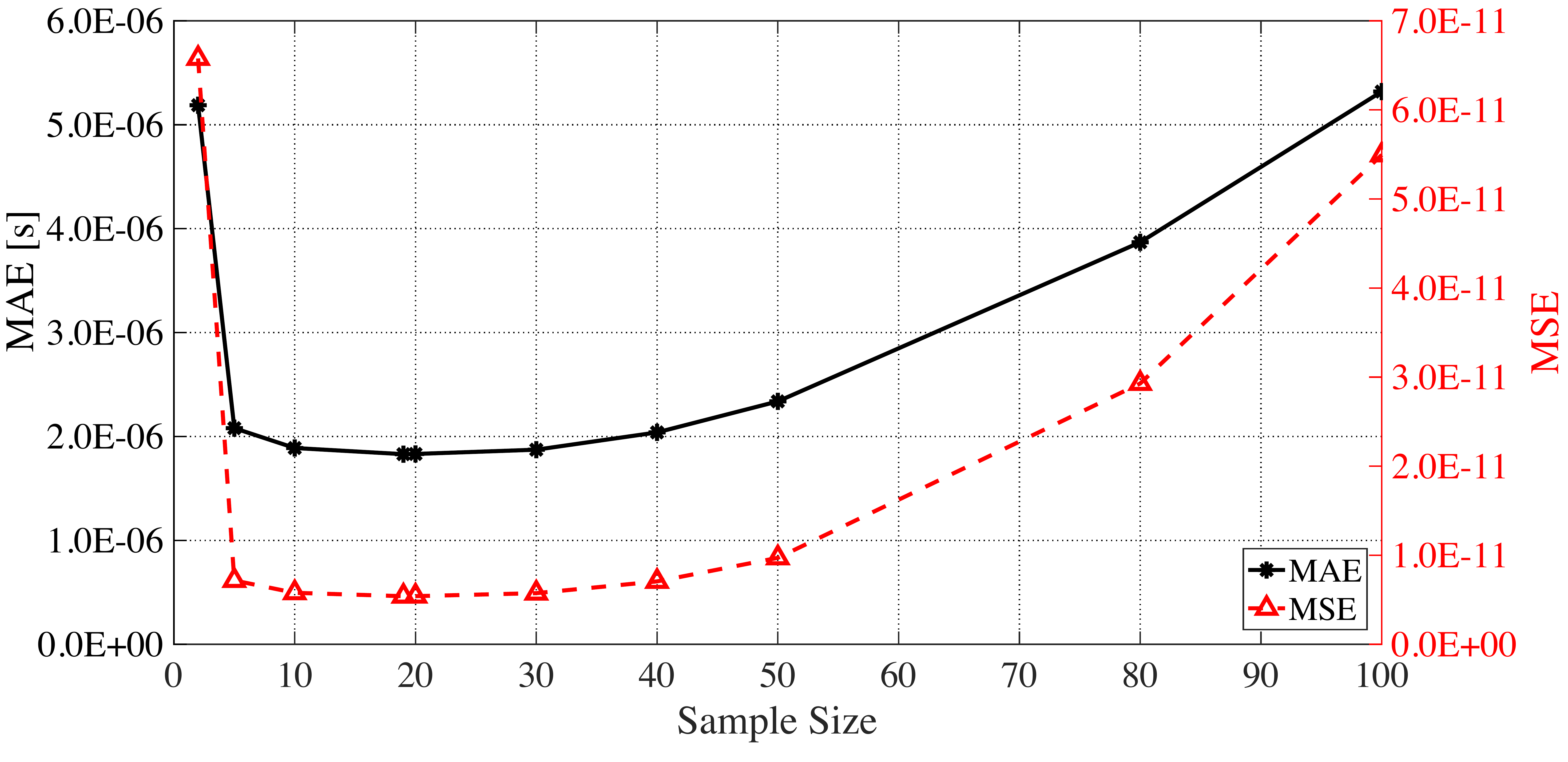} \\
    {\scriptsize (a)} \\
    \includegraphics[width=\linewidth]{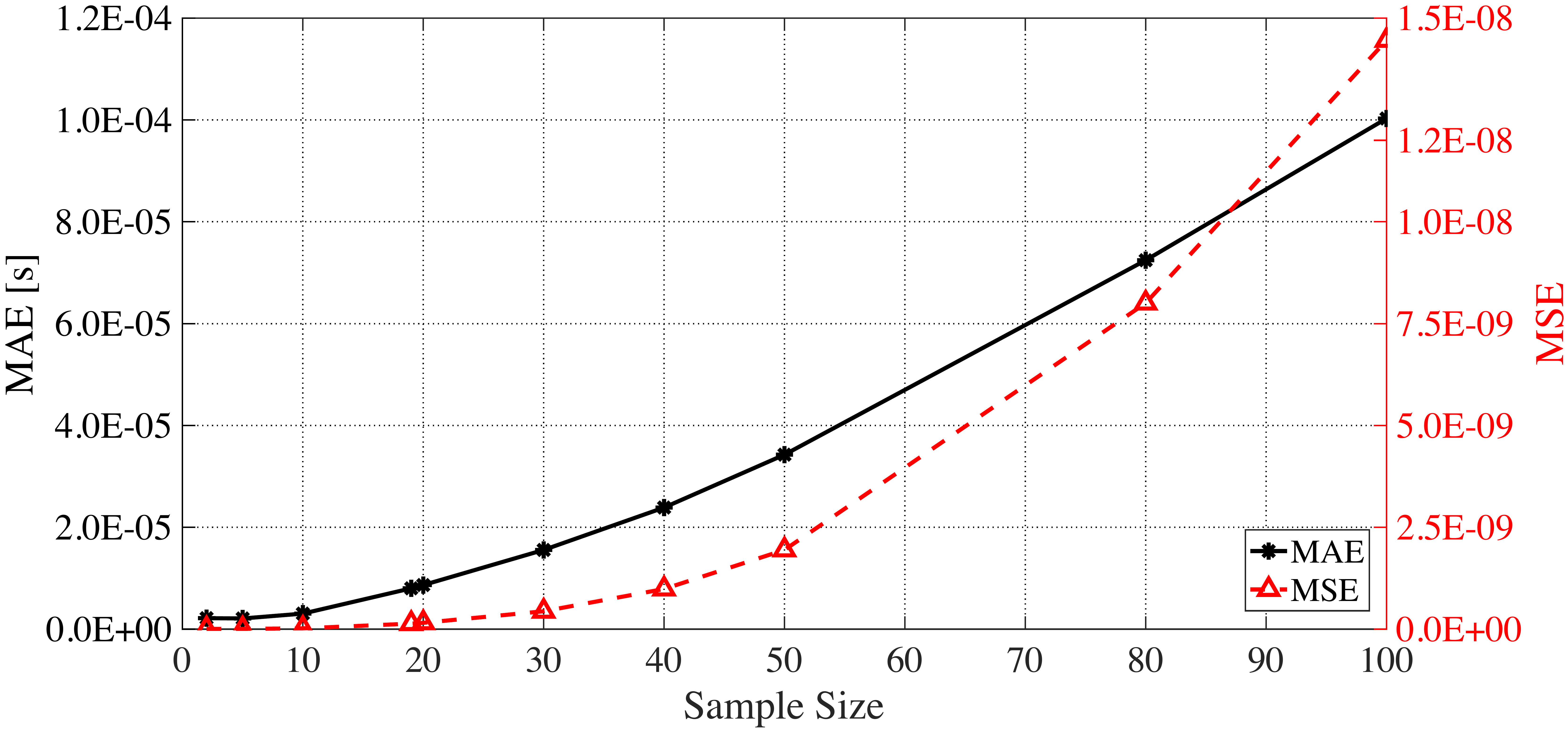} \\
    {\scriptsize (b)} \\
    \includegraphics[width=\linewidth]{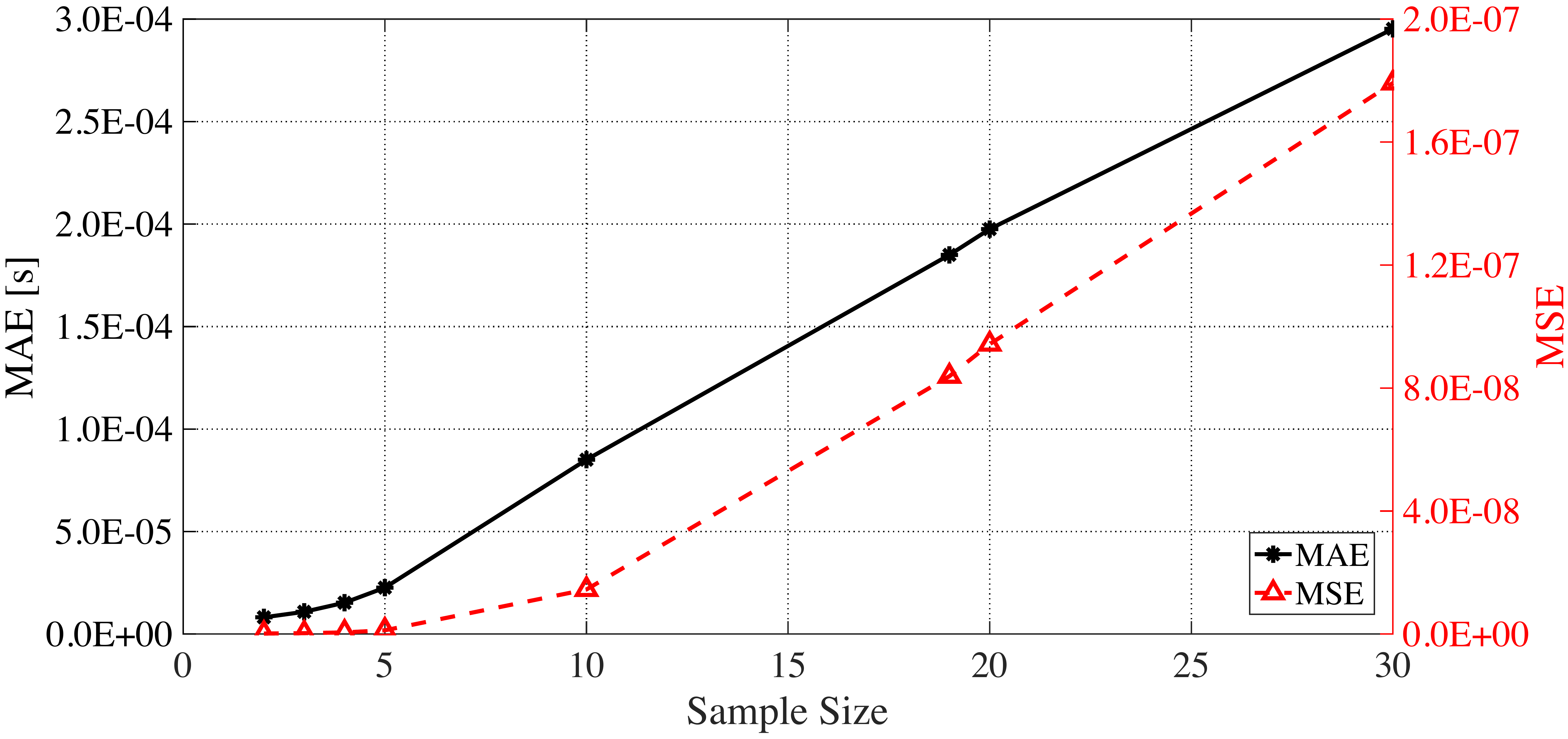} \\
    {\scriptsize (c)}
  \end{center}
  \caption{The effect of sample size on the measurement time estimation of BATS:
    (a) SI=\SI{1}{\s}; (b) SI=\SI{10}{\s} and (c) SI=\SI{100}{\s}.}
  \label{fig:tuning_sample_size}
\end{figure}

Fig.~\ref{fig:tuning_sample_size} shows the effect of sample size on the mean
square error (MSE) and the mean absolute error (MAE) of measurement time
estimation of BATS with different values of SI. From the results, we find that
the MSE and MAE of measurement time estimation are minimal when the sample size
is 19 and 5 for SI of \SI{1}{\s} and \SI{10}{\s}, respectively.
As for the experiments with SI of \SI{100}{\s}, the size of total samples is
quite limited (i.e., only 36 samples from the experiment over \SI{3600}{\s}),
which means, it is quite difficult to apply large value for the aforementioned
sample size. Anyhow, we still tuned the value of sample size in the range of
2--30, and the results showed that, 2 is the best sample size for the experiment
with SI of \SI{100}{\s}.


\begin{figure}[!tb]
  \begin{center}
    \includegraphics[width=\linewidth]{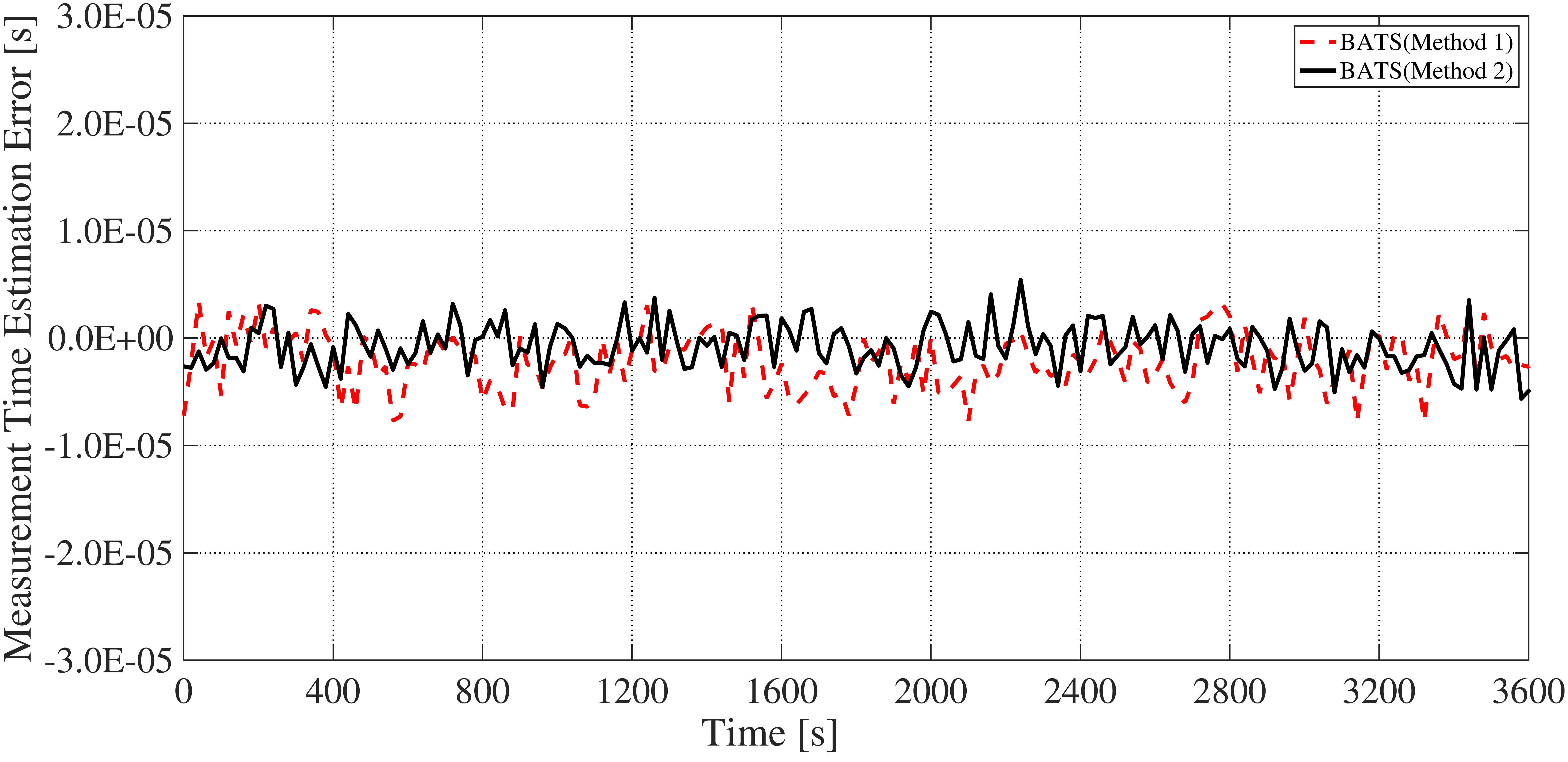} \\
  	{\scriptsize (a)}\\ 
    \includegraphics[width=\linewidth]{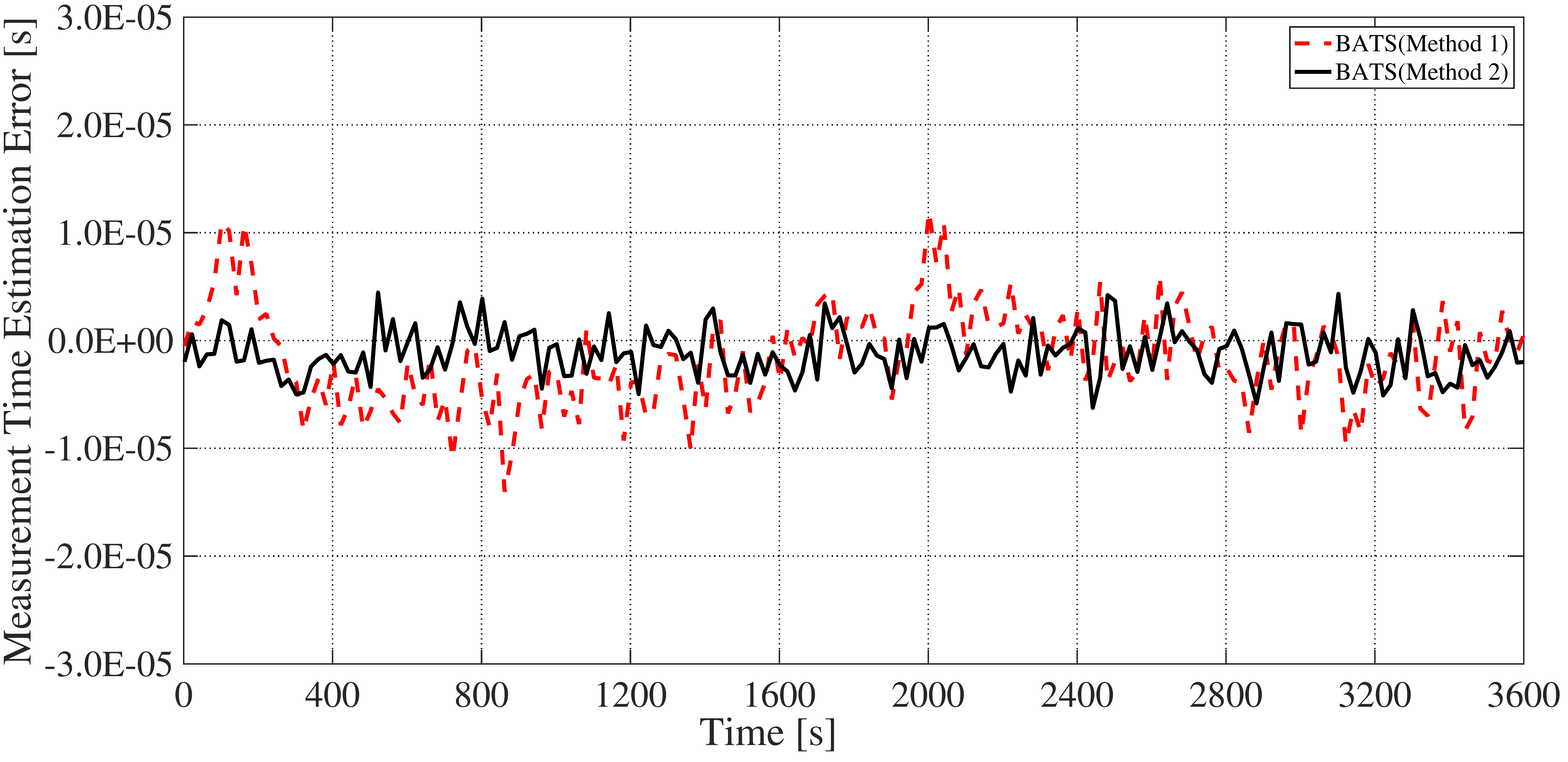} \\
  	{\scriptsize (b)}\\ 
    \includegraphics[width=\linewidth]{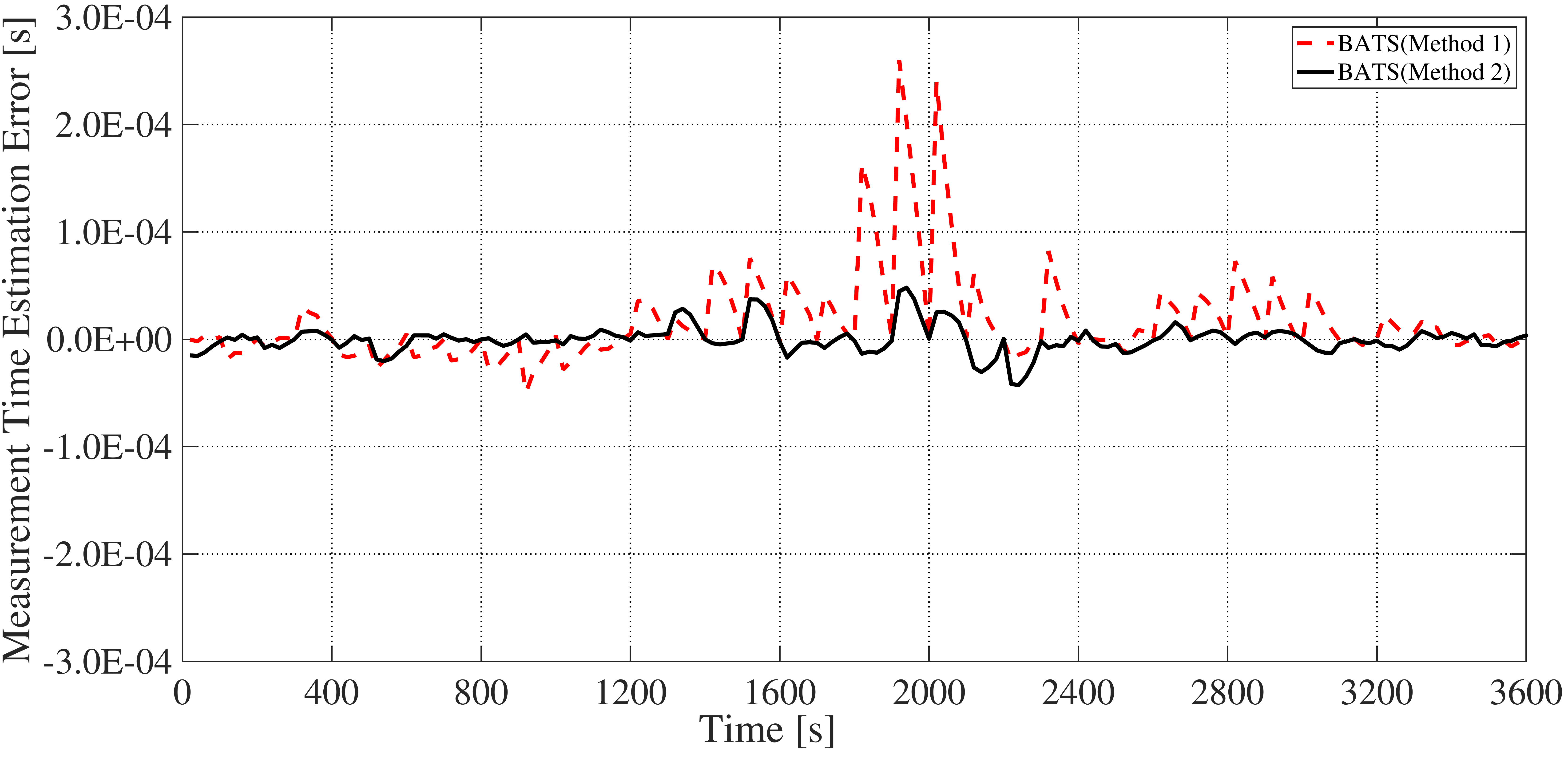} \\
  	{\scriptsize (c)}
  \end{center}
  \caption{Measurement time estimation errors of BATS with ratio-based and
    linear regression methods with SI of (a) \SI{1}{\s}, (b) \SI{10}{\s}, and
    (c) \SI{100}{\s}.}
  \label{fig:MAE_MSEoverTime}
\end{figure}

With the best empirical values of the sample sizes for various values of SI, the
performance of our proposed scheme embedding the linear regression method is demonstrated. The
performance of the method $2$ of linear regression with the best parameter value
outperformed the conventional methods such as the method $1$ based on the
calculation of the cumulative frequency ratio proposed in \cite{rsp} as shown in
Fig.~\ref{fig:MAE_MSEoverTime}. In this figure, the measurement time estimation
errors of our proposed time synchronization scheme based on both traditional
ratio-based method and linear regression method are demonstrated, in which the 
linear regression method outperforms the ratio-based method in all three report
intervals at different levels. In addition, Fig.~\ref{fig:MAE_MSEoverTime}
demonstrated that our proposed reverse asymmetric framework could be employed on
different conventional time synchronization schemes with diverse estimation
methods.

Table~\ref{tab:evaluation_results_sis} summarizes the MAE and MSE of measurement 
time estimation from the experiments during the period of \SI{3600}{\s}. In the 
single-hop scenario, one sensor node is synchronized to one head in the experiments. 
Note that, as the standard FTSP implementation provided in TinyOS library is employed 
in our experiments, so the synchronization accuracy of FTSP is limited to 
millisecond-level\cite{Temperature-Compensated}.
The MAE of measurement time estimation of all three different SIs show that the
proposed scheme with two proposed methods
provides satisfactory precision with minimum of \SI{1.8299}{\us} which is
competitive with consideration to the results presented in other papers---e.g.,
conventional two-way scheme TPSN and one-way schemes FTSP and RSP---which are
also evaluated through real testbeds, however, with drastically fewer message
transmissions as shown in Table.~\ref{tab:evaluation_results_sis}.


In addition, Fig.~\ref{fig:MAE_MSEoverTime} and
Table~\ref{tab:evaluation_results_sis} illustrate that a relatively smaller SI
results in better performance with smaller MSE, which means, the performance of
the proposed scheme with both methods is related to the value of report
intervals. This may result from the drifting of the low-cost crystal oscillator
with up to tens of ppm of clock skew in the resource-constrained sensor node, in
which the drifting range is typically larger in a longer interval.

\subsubsection{Multi-Hop Scenario}
\label{sec:synchronization_performance_multi}
\begin{figure}[!tb]
  \centering
  \includegraphics[width=\linewidth]{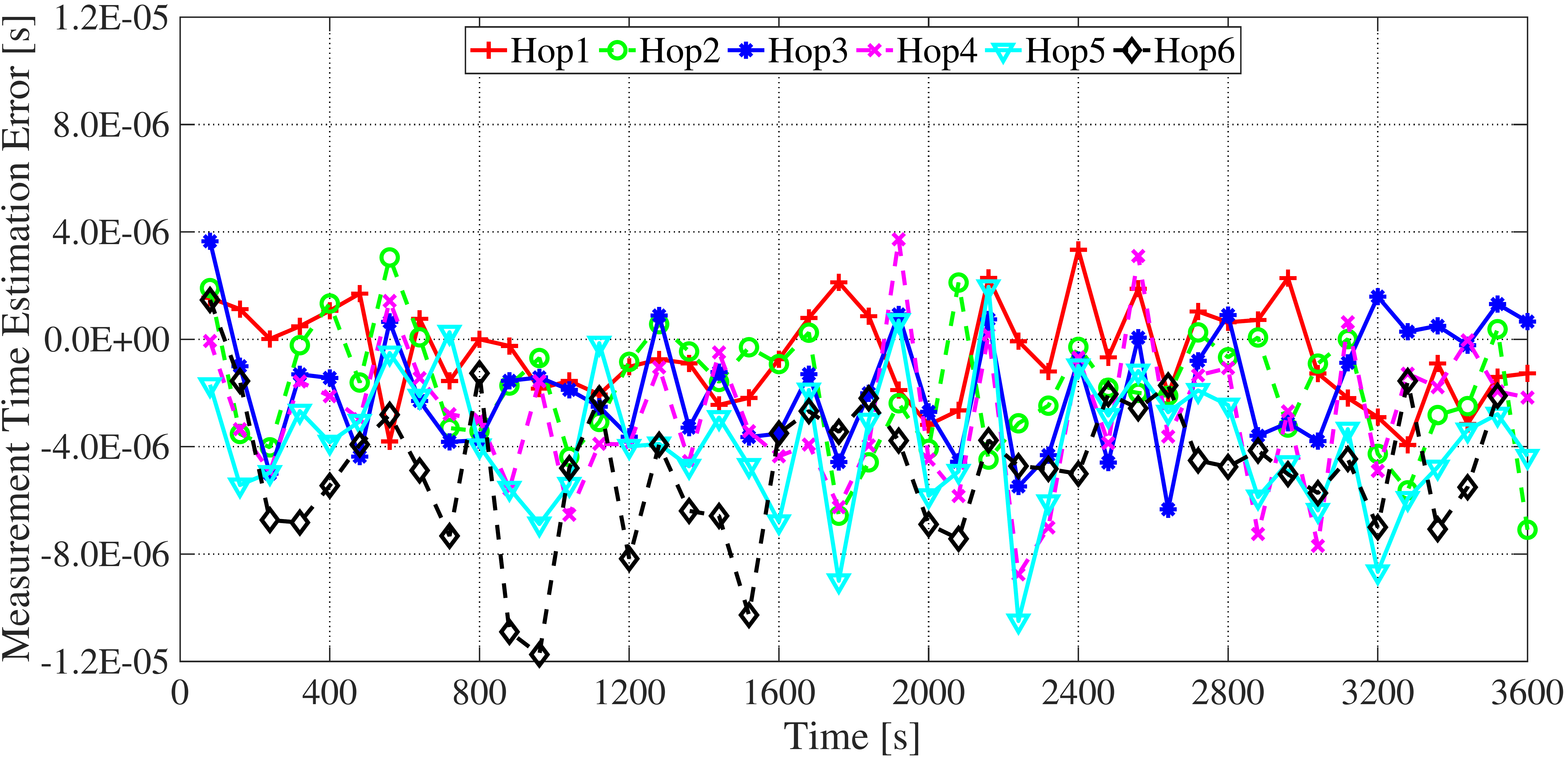}
  \caption{Measurement time estimation errors of BATS for the multi-hop
    scenario.}
  \label{fig:MAEoverTimeonMultihop}
\end{figure}
\begin{figure}[!tb]
  \centering
  \includegraphics[width=\linewidth]{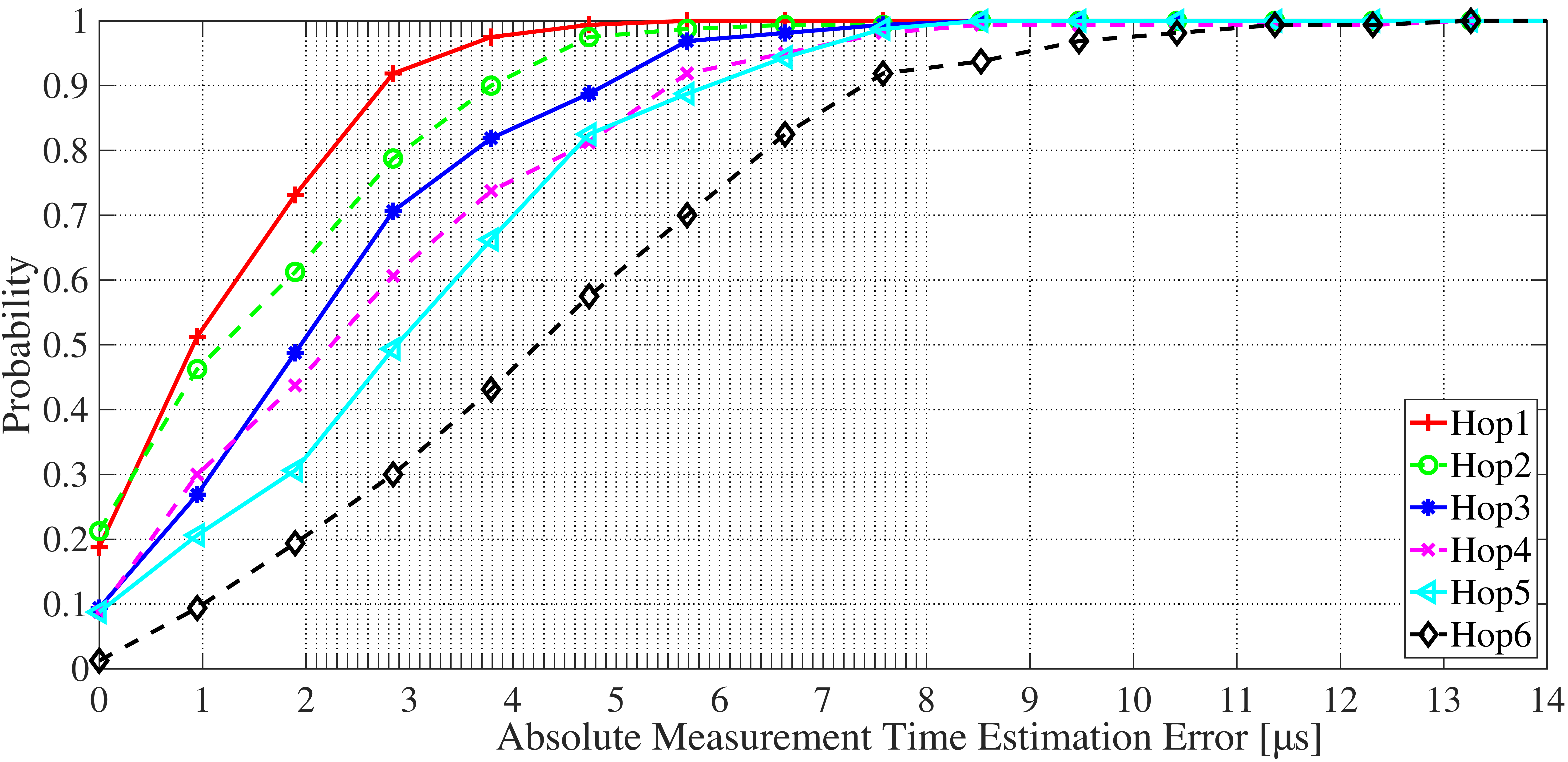}
  \caption{Cumulative distribution functions of the absolute measurement time
    estimation errors of BATS for the multi-hop scenario.}
  \label{fig:Distribution_MAEoverTimeonMultihop}
\end{figure}

\begin{table}[!t]
  \centering
  \begin{threeparttable}
    \caption{MAE and MSE of Measurement Time Estimation of BATS for the
      Multi-Hop Scenario}
    \label{tab:time_sync_results_multihop}
    \setlength{\tabcolsep}{6mm} \centering
    \begin{tabular}{|c|l||r|r|}
      \hline
      \multicolumn{2}{|c||}{Hop Number}
        & \multicolumn{1}{c|}{MAE \tnote{1}}
        & \multicolumn{1}{c|}{MSE \tnote{1}} \\ \hline\hline
      \multirow{6}{*}{Hop}
        & $6$  & 4.2580E-06 & 2.4586E-11 \\ \cline{2-4}
        & $5$  & 3.6149E-06 & 1.8405E-11 \\ \cline{2-4}
        & $4$  & 3.1341E-06 & 1.4519E-11 \\ \cline{2-4}
        & $3$  & 2.4847E-06 & 9.4813E-12 \\ \cline{2-4}
        & $2$  & 1.9455E-06 & 6.0164E-12 \\ \cline{2-4}
        & $1$  & 1.6764E-06 & 4.4735E-12 \\ \hline
    \end{tabular}

    \begin{tablenotes}
    \item[1] Based on the measurement time estimation obtained from
      \SI{3600}{\s} such that the actual performance in real deployment is
      represented.
    \end{tablenotes}
  \end{threeparttable}
\end{table}


Here we investigate the effect of the number of hops on time synchronization
with a flat 6-hop WSN with one head and six sensor nodes. We set the SI to
\SI{1}{\s} and employ the optimal sample size of 19 for the experiment. Note
that we apply the self-data bundling only in order to mainly focus on the effect
of the number of hops, rather than that of bundling, and thereby make the
results more consistent with those of conventional schemes reported in the
literature.

As shown in the Fig.~\ref{fig:MAEoverTimeonMultihop}, the level of fluctuations
of the measurement time estimation errors are roughly proportional to the hop
counts; for instance, the measurement time estimation errors of the node 6 hops
away from the head show the highest fluctuations, while those of the node 1 hop
away from the head show the least fluctuations. This is due to the per-hop
synchronization strategy employed in BATS, where the estimation of the hardware
clock time of a sensor node with respect to the reference clock relies on those
of its upper-layer sensor nodes.

Fig.~\ref{fig:Distribution_MAEoverTimeonMultihop} shows the effect of the number
of hops on time synchronization in a clearer way through the cumulative
distribution functions (CDFs) of \textit{absolute measurement time estimation
  errors}. 90th-percentile absolute measurement time estimation errors for
sensor nodes 1 to 6 hops away from the head are \SI{2.8}{\us}, \SI{3.8}{\us},
\SI{4.9}{\us}, \SI{5.5}{\us}, \SI{5.9}{\us} and \SI{7.4}{\us}, respectively. The
MAEs and MSEs of measurement time estimation are also summarized in
Table~\ref{tab:time_sync_results_multihop}. The results of
Fig.~\ref{fig:Distribution_MAEoverTimeonMultihop} and
Table~\ref{tab:time_sync_results_multihop} demonstrate that the proposed scheme
can provide microsecond-level time synchronization accuracy for all the sensor
nodes in the 6-hop WSN, even though the time synchronization error is cumulative
over the hop count.

\section{Concluding Remarks}
\label{sec:concluding-remarks}
We have proposed BATS, i.e., an energy-efficient time synchronization scheme
based on the framework of reverse asymmetric time synchronization and the reverse
one-way message dissemination; BATS can simultaneously address the major
challenges in WSN time synchronization, i.e., lowering energy consumption and
computational complexity while achieving high time synchronization accuracy.

The major contribution of our work in this paper is three-fold: First, the
reverse asymmetric time synchronization framework is presented. This framework
reassigns the clock parameter estimation procedures from resource-constrained
sensor nodes to the head equipped with abundant computing and power resources,
which leaves only timestamping procedure at the sensor nodes. This reassignment
of time synchronization can not only reduce the computational errors caused by
the limited precision floating-point arithmetic at the sensor nodes but also
bring further potential to use more complex estimation methods at the head,
e.g., those based on machine learning techniques such neural networks as
demonstrated in \cite{NNonTS}; note that, the proposed framework could be employed
by the conventional time synchronization schemes based on both one-way message
dissemination and two-way message exchange, too.

Second, based on the reverse asymmetric time synchronization framework, BATS is
proposed and extended to multi-hop WSNs, which significantly reduces the energy
consumption by eliminating the need of the extra synchronization-related
message---i.e., beacon message and standalone synchronization
message---transmissions.

Third, the actual energy consumption and time synchronization accuracy of BATS
are evaluated through extensive experiments on the real WSN testbed. The results
demonstrate that BATS conserves up to 95\% of the energy consumption by FTSP and
provides \SI{1.8299}{\us} synchronization accuracy in the single-hop
scenario. In case of the multi-hop scenario, the synchronization accuracy for
1-hop and 6-hop sensor nodes are \SI{1.6764}{\us} and \SI{4.2580}{\us}
respectively, which results in \SI{0.5163}{\us} per-hop synchronization error in
average.



As part of BATS, we have also outlined the message bundling procedure, the full
investigation of which would require the consideration of its effect on the
end-to-end delay of the measurement data as well as the time synchronization
performance. In the follow-up work, therefore, we will carry out a systematic
investigation of the effects of the bundling procedure on both synchronization
and delay performance in order to identify potential issues and address them
through more advanced bundling procedures.

Note that the asymmetric time synchronization framework presented in this paper
fits the asymmetric Internet of Things (IoT) deployment which typically consists
of a powerful server and numerous resource-constrained IoT devices
\cite{badihi18:_time_iot}.

\section*{Acknowledgment}
This work was supported by Xi'an Jiaotong-Liverpool University Research
Development Fund (RDF) under grant reference number RDF-16-02-39.

\bibliographystyle{IEEEtran}%
\bibliography{kks}%

\end{document}